\documentclass[aps,pra,superscriptaddress,reprint,showpacs,longbibliography,nofootinbib,amsmath,amssymb]{revtex4-1}
\usepackage[utf8]{inputenc}
\usepackage[T1]{fontenc}

\usepackage{graphicx}
\usepackage{grffile}
\usepackage{microtype}
\usepackage[usenames,dvipsnames]{xcolor}
\usepackage[colorlinks,linkcolor=blue,citecolor=blue,urlcolor=blue]{hyperref}

\def\equationautorefname~#1\null{Eq.~(#1)\null}

\renewcommand{\appendixautorefname}{appendix}
\newcommand{\appref}[1]{\href{#1}{\appendixautorefname~\ref*{#1}}}

\newcommand{\bra}[1]{\langle#1|}
\newcommand{\ket}[1]{|#1\rangle}

\newcommand{\Tn}{\hat{T}_{n}}
\newcommand{\Te}{\hat{T}_{e}}
\newcommand{\Cn}{\hat{C}_{n}}
\newcommand{\Ven}{V_{en}}
\newcommand{\Vnn}{V_{nn}}
\newcommand{\Hm}{\hat{H}_{m}}
\newcommand{\He}{\hat{H}_{e}}
\newcommand{\Hmc}{\hat{H}_{mc}}
\newcommand{\wvib}{\omega_{vib}}
\newcommand{\dR}{\delta\hspace{-0.07em}R}
\newcommand{\DR}{\Delta\hspace{-0.07em}R}

\begin{document}

\title{Cavity-induced modifications of molecular structure in the strong coupling regime}
\author{Javier Galego}
\affiliation{Departamento de F{\'\i}sica Te{\'o}rica de la Materia Condensada and Condensed Matter Physics Center (IFIMAC), Universidad Aut\'onoma de Madrid, E-28049 Madrid, Spain}
\author{Francisco~J.~Garcia-Vidal}
\affiliation{Departamento de F{\'\i}sica Te{\'o}rica de la Materia Condensada and Condensed Matter Physics Center (IFIMAC), Universidad Aut\'onoma de Madrid, E-28049 Madrid, Spain}
\affiliation{Donostia International Physics Center (DIPC), E-20018 Donostia/San Sebastian, Spain}
\author{Johannes Feist}
\email{johannes.feist@uam.es}
\affiliation{Departamento de F{\'\i}sica Te{\'o}rica de la Materia Condensada and Condensed Matter Physics Center (IFIMAC), Universidad Aut\'onoma de Madrid, E-28049 Madrid, Spain}

\begin{abstract}
In most theoretical descriptions of collective strong coupling of organic molecules to a cavity mode, the molecules are modeled as simple two-level systems. This picture fails to describe the rich structure provided by their internal rovibrational (nuclear) degrees of freedom. We investigate a first-principles model that fully takes into account both electronic and nuclear degrees of freedom, allowing an exploration of the phenomenon of strong coupling from an entirely new perspective. First, we demonstrate the limitations of applicability of the Born-Oppenheimer approximation in strongly coupled molecule-cavity structures. For the case of two molecules, we also show how dark states, which within the two-level picture are effectively decoupled from the cavity, are indeed affected by the formation of collective strong coupling. Finally, we discuss ground-state modifications in the ultra-strong coupling regime and show that some molecular observables are affected by the collective coupling strength, while others only depend on the single-molecule coupling constant.
\end{abstract}

\pacs{%42.50.Nn, %quantum optical phenomena
71.36.+c, % Polaritons
78.66.Qn, % Polymers; organic compounds
%82.20.-w, % Chemical kinetics and dynamics
82.20.Kh, % Potential energy surfaces for chemical reactions
%71.35.-y, % Excitons and related phenomena
73.20.Mf, % Collective excitations (including excitons, polarons, plasmons and other charge-density excitations)
}
\date{\today}
\maketitle

\section{Introduction}

Strong coupling in quantum electrodynamics is a well-known phenomenon that occurs when the coherent energy exchange between a light mode and quantum emitters is faster than the decay and decoherence of either constituent~\cite{Thompson1992,Weisbuch1992}. The excitations of the system are then hybrid light-matter excitations, so-called polaritons, that combine the properties of both constituents. Exploiting these properties enables new applications such as polariton condensation under collective strong coupling to excitons (excited electron-hole pairs) in semiconductors~\cite{Kasprzak2006,Balili2007} and organic materials~\cite{Kena-Cohen2010,Plumhof2014,Daskalakis2014}. Organic materials present a particularly favorable case, as the Frenkel excitons in these materials possess large binding energies, large dipole moments, and can reach high densities. This enables Rabi splittings $\Omega_R$ (the energy splitting between the polaritons) up to more than $1$~eV~\cite{Lidzey1998,Schwartz2011,Kena-Cohen2013}, a significant fraction of the uncoupled transition energy. These properties allow for strong coupling to many kinds of electromagnetic (EM) modes~\cite{Torma2015}, such as cavity photons~\cite{Lidzey1998,Schwartz2011,Kena-Cohen2008}, surface plasmon polaritons~\cite{Bellessa2004,Dintinger2005,Hakala2009,Vasa2010}, surface lattice resonances~\cite{Rodriguez2013,Vakevainen2014}, or localized surface plasmons~\cite{Baudrion2013,Zengin2015}.

\begin{figure}
\includegraphics[width=\linewidth]{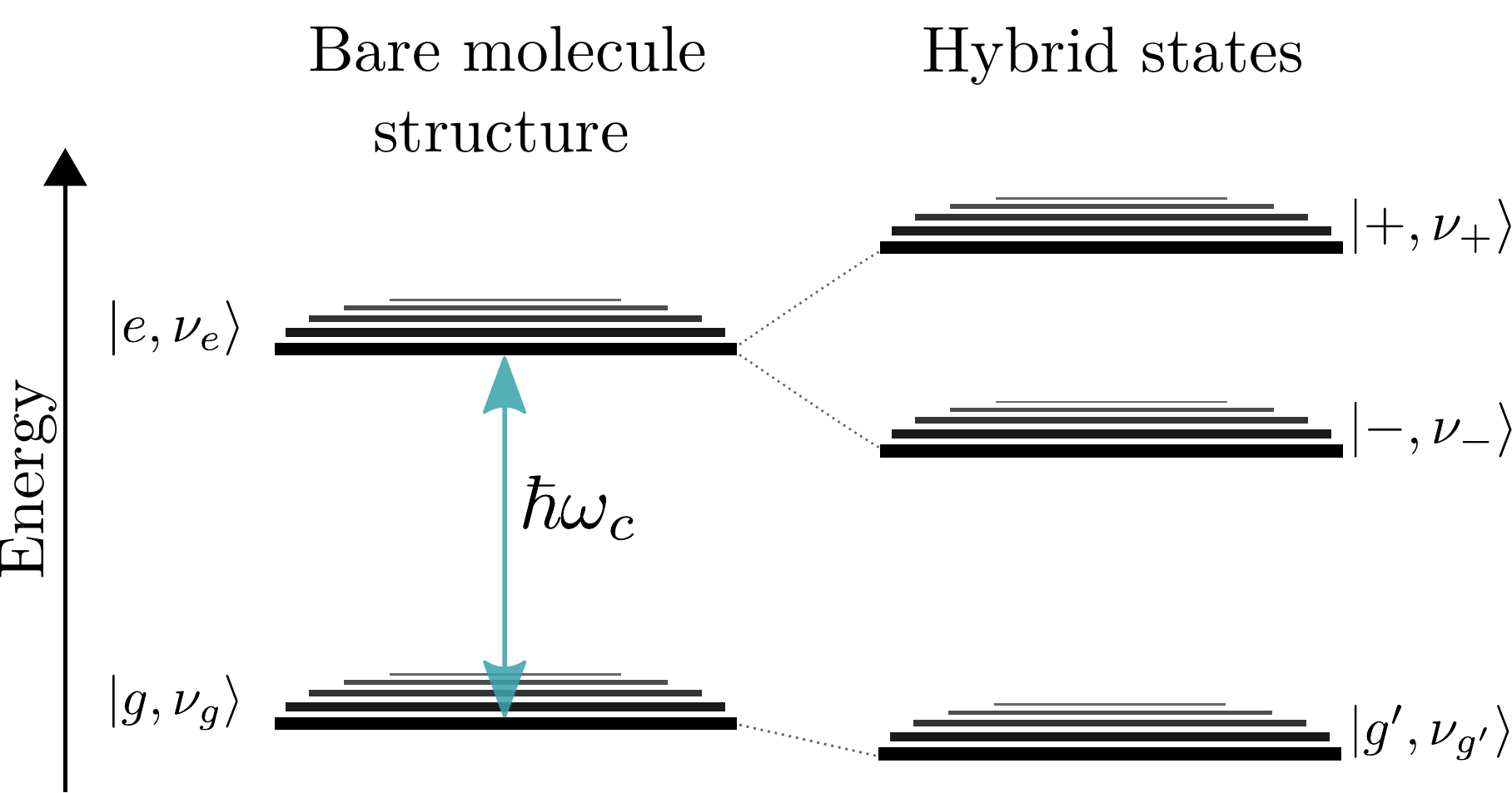}
\caption{Illustration of energy level structure of a bare complex molecule and the hybrid states that result in the strong coupling regime with a photonic mode of energy $\hbar\omega_c$, resonant with the molecular excitation.}
\label{fig:level_structure}
\end{figure}

While organic molecules are thus uniquely suited to achieving strong coupling, they are not simple two-level quantum emitters, but rather have a complicated level structure including not only electronic excitations, but also rovibrational degrees of freedom (schematically depicted in \autoref{fig:level_structure}). It has been experimentally demonstrated that strong coupling can modify this structure, in the sense that material properties and chemical reaction rates change~\cite{Hutchison2012,Hutchison2013,Wang2014b}. However, the models used to describe strong coupling are often focused on macroscopic descriptions~\cite{Carusotto2013}, and most microscopic models do treat organic molecules as two-level systems (see \cite{Michetti2015} for a recent review). When the rovibrational degrees of freedom are taken into account, this is often done using effective decay and dephasing rates~\cite{Gonzalez-Tudela2013}, with a few works explicitly including a phononic degree of freedom~\cite{Mazza2013,Cwik2014}. All of these approaches only provide limited insight into the effects of strong coupling on molecular structure.

In the present work, we thus aim for a microscopic description of strong coupling with organic molecules. We introduce a simple first-principles model that fully describes nuclear, electronic and photonic degrees of freedom, but can be solved without approximations. This allows us to provide a simple picture for understanding the induced modification of molecular structure. %We will focus on a number of fundamental questions:

In \autoref{sec:single_molecule}, after introducing the model, we discuss under which conditions and in which form the Born-Oppenheimer approximation (BOA)~\cite{Born1927,Tully2000} is valid in the strong coupling regime for a single molecule. The BOA is widely used in molecular and solid state physics and quantum chemistry, and provides a simple picture of nuclei moving on effective potential energy surfaces (PES) generated by the electrons, which underlies most of the current understanding of chemical reactions~\cite{Tully2000}. However, the BOA depends on the separation of electronic and nuclear energy scales, i.e., the fact that electrons typically move much faster than nuclei. It could thus conceivably break down when an additional, intermediate timescale is introduced under strong coupling to an EM mode. The speed of energy exchange between field and molecules is determined by the Rabi frequency $\Omega_R$, and typical experimental values of hundreds of meV land squarely between typical nuclear ($\simeq100~$meV) and electronic ($\simeq2$~eV) energies. We show that the BOA indeed breaks down at intermediate Rabi splittings, but remains valid when $\Omega_R$ becomes large enough. For cases where it breaks down, we show that the non-BO coupling terms can be obtained to a good approximation without requiring knowledge of the electronic wave functions.

In \autoref{sec:two_molecules}, we focus on the effects of strong coupling when more than one molecule is involved, using two molecules as the simplest test case. In experiments, strong coupling is achieved by collective coupling to a large number of molecules, under which the Rabi frequency is enhanced by a factor of $\sqrt{N}$. The number of emitters $N$ is on the order of $\gtrsim10^5$ within cavities~\cite{Lidzey1998,Schwartz2011,Kena-Cohen2008,Kena-Cohen2013}, with plasmonic nanoparticle modes allowing to reduce this to $N\sim100$~\cite{Zengin2015}. In this context, it is well known that only a small fraction of the collective electronic excitations are strongly coupled~\cite{Houdre1996,Agranovich2011,Michetti2015}, with a large number of ``dark'' or ``uncoupled'' modes that show no mixing with the EM mode and no energy shift. We show that even these dark modes are affected by strong coupling, with the nuclear motion of separated molecules becoming correlated.

In \autoref{sec:ultrastrong_coupling}, we focus on the so-called \emph{ultrastrong} coupling regime, where the Rabi frequency reaches a significant fraction of the electronic transition energy, as achieved in experiments. In this regime, not just excited-state, but also ground-state properties are modified---for example, the ground state acquires a photonic contribution~\cite{DeLiberato2007,Carusotto2013}. Accordingly, we discuss whether ground-state chemical properties of organic molecules could be modified by strong coupling. This also allows us to partially answer the open question what strong coupling means for modifications of chemical structure~\cite{George2015}, i.e., whether ``all'' molecules are modified by it, or only a small subset, or whether we necessarily have to invoke collective modes even when discussing ``single-molecule'' effects. We show that while some observables, such as energy shifts, are determined by the \emph{collective} Rabi frequency, but other observables, such as the shift in ground-state bond length, are instead determined by the \emph{single-molecule} coupling strength $\propto \Omega_R/\sqrt{N}$.

For simplicity, we only treat a single EM mode and completely neglect dissipation in the following. We use atomic units unless stated otherwise ($4\pi\varepsilon_0=\hbar=m_e=e=1$, with electron mass $m_e$ and elementary charge $e$).

\section{Single Molecule}\label{sec:single_molecule}
In this section, we introduce our model for a single molecule coupled to an EM mode. Due to the exponential scaling with the degrees of freedom, solving the full time-independent Schrödinger equation for an organic molecule without the BOA is an extremely challenging task that even modern supercomputers can only handle for very small molecules. We thus employ a reduced-dimensionality model that we can easily solve, both for the bare molecule and after coupling to an EM mode.

\subsection{Method}
\subsubsection{Bare molecule}

We work within the single-active-electron approximation (SAE), in which all but one electron are frozen around the nuclei, and additionally restrict the motion of the active electron to one dimension, $x$. Furthermore, we only treat one nuclear degree of freedom, the reaction coordinate $R$. This could correspond to the movement of a single bond in a molecule, but can equally well represent collective motion, e.g., the breathing mode of a carbon ring. The effective molecular Hamiltonian then highly resembles that of a one-dimensional diatomic molecule,
\begin{equation}
\Hm = \Tn + \Te + \Ven(x;R) + \Vnn(R),
\label{eq:bare_hamiltonian}
\end{equation}
where $\Tn = \frac{\hat{P}^2}{2M}$ and $\Te = \frac{\hat{p}^2}{2}$ are the nuclear and electronic kinetic energy operators (with $\hat{P}$, $\hat{p}$ the corresponding momenta), and $M$ is the nuclear mass. The potentials $\Ven(x,R)$ and $\Vnn(R)$ represent the effective electron-nuclei and internuclear interactions, where we assume two nuclei located at $x=\pm R/2$. These potentials encode the information about the frozen electrons as well as the nuclear structure of the molecule, and can be adjusted to approximately represent different molecules.

The electron-nucleus interaction $\Ven$ contains the interaction of the active electron with each nucleus, as well as with the frozen electrons surrounding it. Assuming a nuclear charge of $Z$, we have $2Z-1$ frozen electrons distributed across the two nuclei. For large distances, the active electron should thus feel a Coulomb potential with an effective charge of $\frac12$ from each nucleus. Conversely, at very small distances, the active electron is not affected by the cloud of frozen electrons and feels an effective charge of $Z$. Since we are working within one dimension, we use a soft Coulomb potential to take into account that the electron avoids the singularity at the nucleus. We choose a simple model potential fulfilling these conditions:
\begin{equation}
\Ven(r)= -\frac{\frac{1}{2}+(Z-\frac{1}{2})e^{-\frac{r}{r_0}}}{\sqrt{r^2 +\alpha ^2}} ,
\end{equation}
where $\alpha$ is the softening parameter, $r_0$ describes the localization of the frozen electrons around the nucleus, and $r$ is the electron-nucleus distance. The total potential is thus $\Ven(x,R)=\Ven(|x-R/2|) + \Ven(|x+R/2|)$.

The internuclear potential $\Vnn(R)$ represents the interaction between the nuclei and the $2Z-1$ frozen electrons, i.e., the ground state potential energy surface of the molecular ion. We model this surface by a Morse potential
\begin{equation}
\Vnn(R)=D_e \left(1-e^{A(R-R_0)}\right)^2,
\end{equation}
which adds three new parameters: the dissociation energy $D_e$, the equilibrium distance $R_0$ and the width of the potential well $A$. By tuning the seven free parameters we have at our disposal ($M$, $Z$, $\alpha$, $r_0$, $D_e$, $R_0$ and $A$), we can approximately fit both the electronic and vibrational structure and absorption spectrum to those of real organic molecules.

We can now solve the stationary Schrödinger equation $\Hm\Psi(x,R) =E\Psi(x,R)$ for the bare-molecule Hamiltonian \autoref{eq:bare_hamiltonian} without further approximations by representing $\Hm$ on a two-dimensional grid in $x$ and $R$. For a bare molecule, the results are virtually identical to those obtained within the BOA and thus not shown here.

We next give a short description of the Born-Oppenheimer approximation for completeness (see~\cite{Born1927,Tully2000} for more details). As mentioned above, the basic idea is to exploit the separation between nuclear and electronic timescales and to assume that the electrons perfectly follow nuclear rearrangements without changing state (i.e., adiabatically). This is achieved by separating the Hamiltonian into the nuclear kinetic energy $\Tn$ and an electronic Hamiltonian $\He(x;R) = \Hm(x,R)-\Tn$ that only depends on $R$ parametrically. Diagonalizing $\He$ yields a set $\{\phi_k\}$ of electronic eigenstates for every $R$, with $\He(x;R) \phi_k(x;R) = E_k(R) \phi_k(x;R)$. Without loss of generality, each total eigenstate $\Psi^i$ can be represented by $\Psi^i(x,R) = \sum_k \phi_k(x;R) \chi^i_k(R)$. Inserting this expansion into the Hamiltonian \autoref{eq:bare_hamiltonian} leads to a set of coupled differential equations,
\begin{equation}
(\Tn + E_k) \chi^i_k(R) + \sum_{k'} \Cn^{kk'} \chi^i_{k'} = E \chi^i_k(R) ,
\label{eq:BOA_coupled}
\end{equation}
with nuclear motion taking place on potential energy surfaces (PES) $E_k(R)$ that are coupled through correction terms $\Cn^{kk'} = \bra{\phi_k} \Tn \ket{\phi_{k'}}_x + \bra{\phi_k} \frac{\hat{P}}{2M} \ket{\phi_{k'}}_x \hat{P}$, where the subscript $x$ indicates that the integration in the brakets is only over the electronic coordinate. The Born-Oppenheimer approximation now consists in neglecting the intersurface couplings $\Cn^{kk'}$, which can be shown to be small when the electronic levels are well-separated. This gives a set of \emph{independent} PES $E_k(R)$ on which the nuclei move, where each eigenstate is a product of a single electronic and nuclear wave function, $\Psi^i_k(x,R) = \phi_k(x;R) \chi^i_k(R)$. The different nuclear functions on each electronic curve correspond to rotational or vibrational excitation. This picture of nuclear motion on PES is extremely powerful and underlies most of the current understanding of chemical reactions~\cite{Tully2000}. The question of its validity in the strong coupling regime is thus of central importance for the possible modification of chemical reactions and structure through strong coupling.

\begin{figure}
\includegraphics[width=\linewidth]{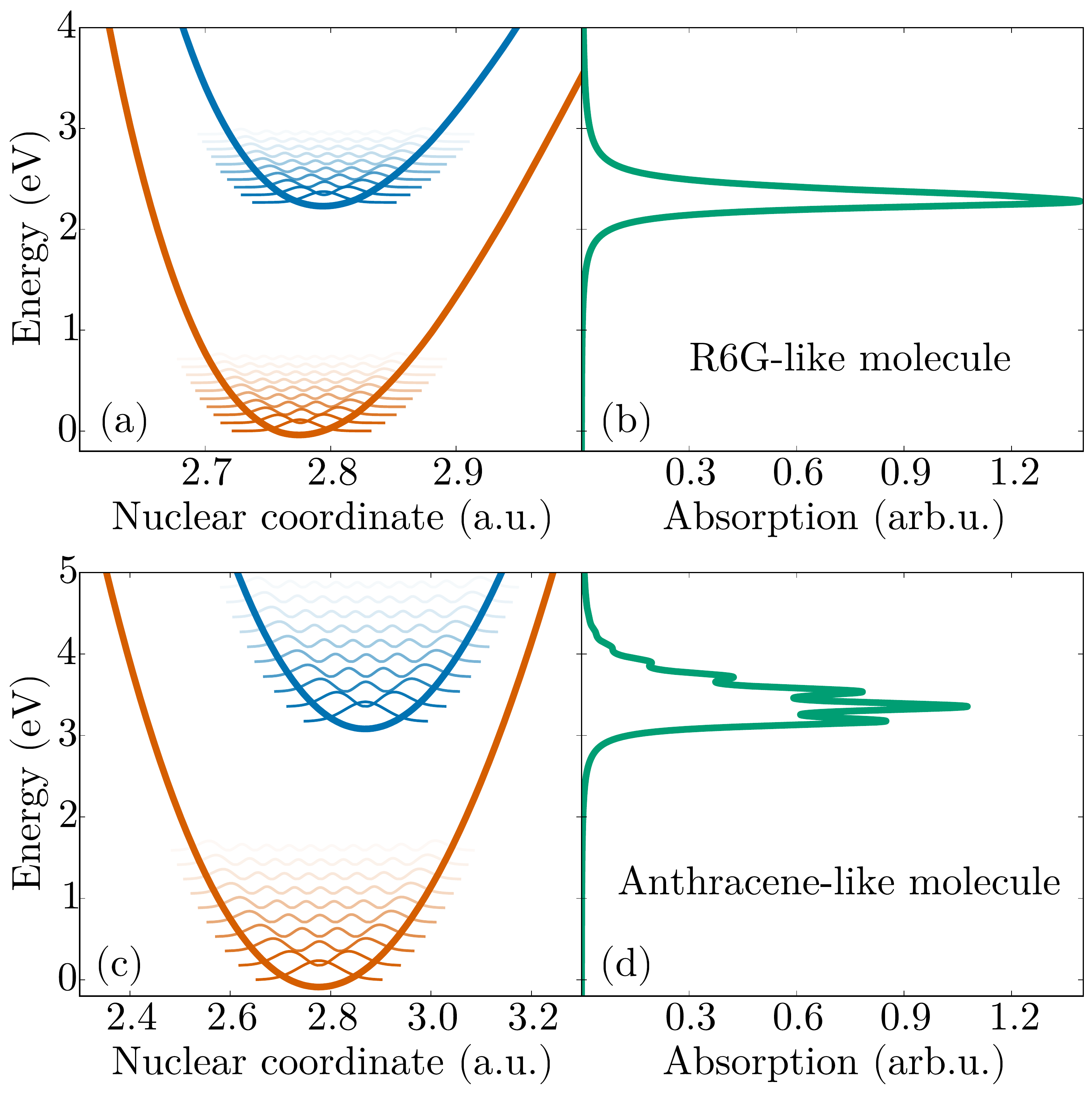}
\caption{Bare-molecule potential energy surfaces of the two first electronic states in the BOA for (a) the rhodamine 6G-like model molecule and (c) the anthracene-like model molecule. The vibrational levels and associated nuclear probability densities are represented on top of the PES. (b) and (d): Absorption spectrum for the (b) R6G-like and (d) anthracene-like molecule in arbitrary units.}
\label{fig:structure_molAB}
\end{figure}

In the following, we will focus on two model molecules, which approximately reproduce the absorption spectra of rhodamine 6G (R6G) and anthracene molecules that are commonly used in experimental realizations of strong coupling~\cite{Kena-Cohen2008,Hakala2009,Rodriguez2013}. 
Only the first two PES, corresponding to the ground $E_g(R)$ and first electronically excited $E_e(R)$ state, play a role in the results discussed in the following. They are shown in \autoref{fig:structure_molAB}a and \autoref{fig:structure_molAB}c, together with the nuclear probability densities $|\chi(R)|^2$ for the lowest vibrational levels on each PES. Importantly, the two models differ significantly in two relevant quantities: the vibrational mode frequency $\wvib$ and the offset $\DR$, i.e., the change in equilibrium distance between the ground and excited PES. This offset is related to the strength of the electron-phonon interaction and influences the Stokes shift between emission and absorption~\cite{May2011}. The R6G-like model has relatively small vibrational spacing $\wvib\approx70$~meV and small offset $\DR\approx0.018$~a.u., while the anthracene-like model has large vibrational spacing $\wvib\approx180$~meV and large offset $\DR\approx0.092$~a.u.. Accordingly, their absorption spectra (\autoref{fig:structure_molAB}b and \autoref{fig:structure_molAB}d, obtained from \autoref{eq:absorption} below) are qualitatively different, with anthracene showing a broader absorption peak with well-resolved vibronic subpeaks.

\subsubsection{Molecule-photon coupling}
We now add a photonic mode and its coupling to the molecule (within the dipole approximation) into the molecular Hamiltonian,
\begin{equation}
\Hmc = \Hm + \omega_c \hat{a}^{\dagger} \hat{a} + g\hat{\mu}(\hat{a}^{\dagger}+\hat{a}),
\label{eq:coupling_hamiltonian}
\end{equation}
where $\hat{\mu}$ is the dipole operator of the molecule ($\hat\mu = x$ in our case), $\hat{a}^\dagger$ and $\hat{a}$ are the creation and annihilation operators for the bosonic EM field mode, $\omega_c$ is its frequency, and $g$ is the coupling strength constant, given by the electric field amplitude (along the $x$-axis) of a single photon. In the following, we always set the photon energy $\omega_c$ to achieve ``zero detuning'', with $\omega_c$ at the absorption maximum of the molecule. This gives $\omega_c \approx E_e(R_e)-E_g(R_e)$, where $R_e$ is the equilibrium position at which $E_g(R)$ has its minimum.

To provide some context for the field strengths used in the following, we note that for a typical microcavity with a mode volume $V\approx\lambda_c^3$, one obtains $g=\sqrt{\frac{\hbar\omega_c}{2\varepsilon_0 V}} \approx 1.34\times 10^{-7} \omega_c^2$~a.u.\ (for $\omega_c$ given in eV). However, for an effective mode volume close to the current record achieved in plasmonic nano-antennas, $V\approx1.3\times 10^{-7}\lambda_c^3$~\cite{Kim2015}, the single-particle coupling reaches $g\approx 3.72\times 10^{-4} \omega_c^2$~a.u.\ ($\omega_c$ again in eV). We furthermore note that the ground-to-excited state dipole transition moments of our model molecules are on the order of $1~\mathrm{a.u.}\approx2.54$~D, i.e., almost an order of magnitude smaller than in typical organic molecules~\cite{Moll1995}.

\begin{figure}
\includegraphics[width=\linewidth]{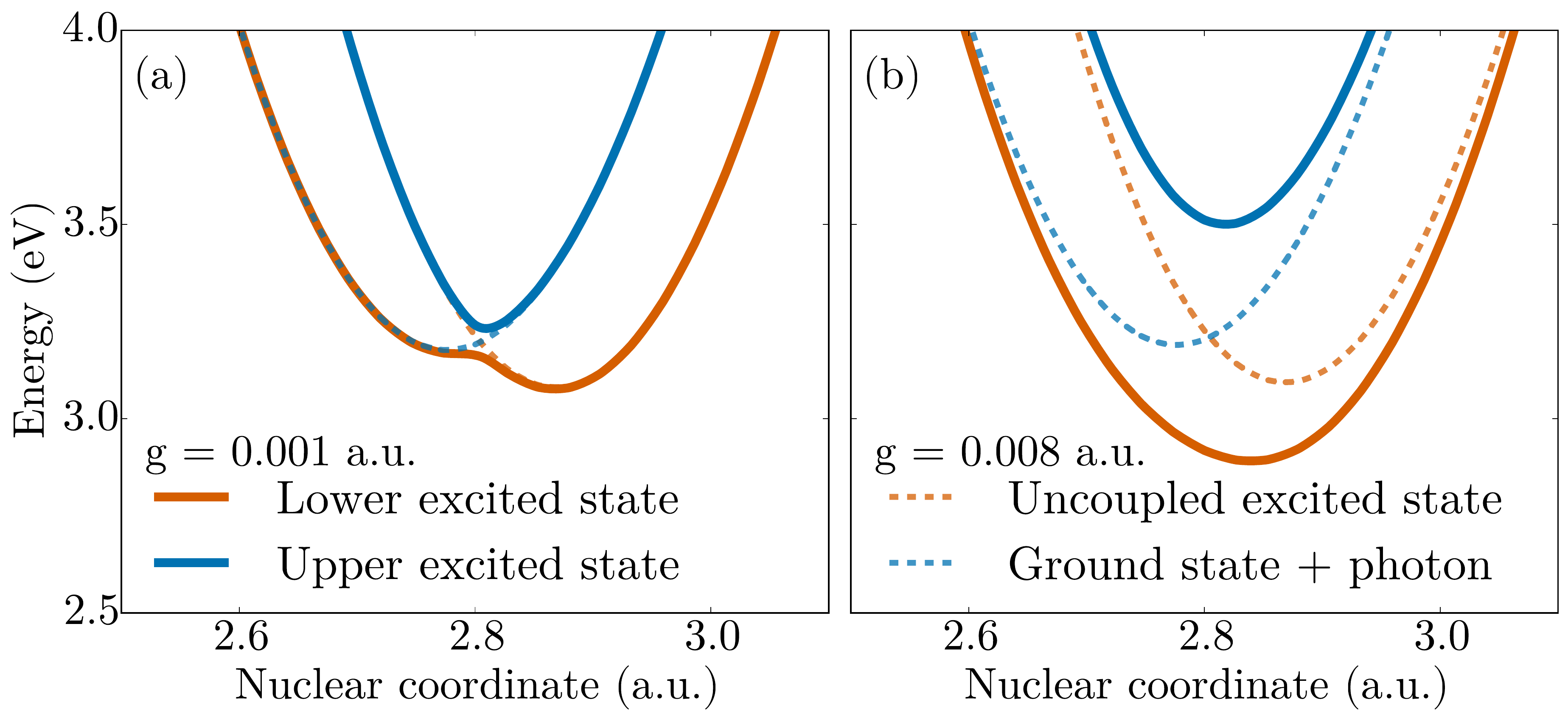}
\caption{Strongly coupled electronic PES (solid lines) in the singly excited subspace, for the anthracene-like molecule for (a) $g=0.001$ a.u.\ and (b) $g=0.008$ a.u.. The dashed lines show the corresponding uncoupled states: A molecule in the first excited state, $E_e(R)$, and a molecule in the ground state with one photon present, $E_g(R)+\omega_c$.}
\label{fig:anticrossing_molB}
\end{figure}

Compared to the bare-molecule case, the Hamiltonian now includes a new degree of freedom, the photon number $n\in\{0,1,2,\ldots\}$, with the system eigenstates defined by $\Hmc\Psi(x,n,R) = E \Psi(x,n,R)$. As discussed above, the typical energies associated with strong coupling in organic molecules are somewhere between the nuclear and electronic energies. A priori, this suggests two options of performing the BOA: the additional terms introduced by the photonic degree of freedom could be grouped either with the ``slow'' nuclear motion or with the ``fast'' electronic Hamiltonian. However, as the photon couples to the electron, grouping it with the nuclear terms necessarily leads to additional off-diagonal terms in \autoref{eq:BOA_coupled}, and no independent PES on which the nuclei move could be obtained. Consequently, the only way to maintain the usefulness of the BOA is to include the photonic degree of freedom within the electronic Hamiltonian, leading to a new set of ``strongly coupled PES''.

We first focus on the singly excited subspace, within which the splitting between polaritons is observed. Here, either the molecule is electronically excited and no photons are present, or the molecule is in its electronic ground state and the photon mode is singly occupied. At zero coupling ($g=0$), this gives two uncoupled PES ($E_e(R)$ and $E_g(R)+\omega_c$, dashed curves in \autoref{fig:anticrossing_molB}) that cross close to $R_e$ for our choice of $\omega_c$. When the electron-photon coupling is non-zero but small, a narrow avoided crossing develops instead (solid lines in \autoref{fig:anticrossing_molB}a), while for large coupling strengths, the energy exchange between photonic and electronic degrees of freedom is so fast that we observe two entirely new PES (\autoref{fig:anticrossing_molB}b), which can not be easily associated with either of the uncoupled PES. Instead, they become hybrid polaritonic PES that contain a mixture of electronic and photonic excitation, the hallmark of the strong coupling regime.

As discussed above, the BOA is known to be valid when two PES are sufficiently separated from each other. This implies that the BOA breaks down when $g$ is small and the two PES possess a narrow avoided crossing. This in itself is not a surprising result---when the electron-photon coupling is very small, the system is not even in the strong coupling regime, and the photon mode is better treated as a small perturbation. Fortunately, the weak coupling regime is also not interesting from the standpoint of understanding or modifying molecular structure through strong coupling. The real question thus must be: \emph{How strong} does the electron-photon coupling have to be for the BOA to be valid, and is this condition fulfilled for realistic experimental parameters? In order to better quantify the agreement between the BOA and the full solution, we next turn to an easily measured physical observable: the absorption spectrum.

\subsection{Absorption}
The absorption cross section at frequency $\omega$ can be calculated from the polarizability as~\cite{Rescigno1975,BonKre1997}
\begin{equation}
\sigma(\omega) = \frac{4\pi\omega}{c} \operatorname{Im} \lim_{\varepsilon\to0} \sum_k \frac{|\bra{\psi_k}\hat{\mu}\ket{\psi_0}|^2}{\omega_k-\omega_0 - \omega - i\varepsilon} ,
\label{eq:absorption}
\end{equation}
where the sum runs over all eigenstates $\ket{\psi_k}$ of the system (with energies $\omega_k$) and $\ket{\psi_0}$ is the ground state. As we do not include incoherent processes in our calculation, this would give $\delta$-like peaks in the absorption cross section. In the plots shown in the following, we instead introduce a phenomonological width representing losses and pure dephasing by setting $\varepsilon$ to a small non-zero value, such that the absorption cross section becomes a sum of Lorentzians. For the bare-molecule case without coupling to an EM mode, the absorption spectra of our two model molecules approximately agree with those of R6G (\autoref{fig:structure_molAB}b, \cite{Rodriguez2013}) and anthracene (\autoref{fig:structure_molAB}d, \cite{Kena-Cohen2008}).

\begin{figure}
\includegraphics[width=\linewidth]{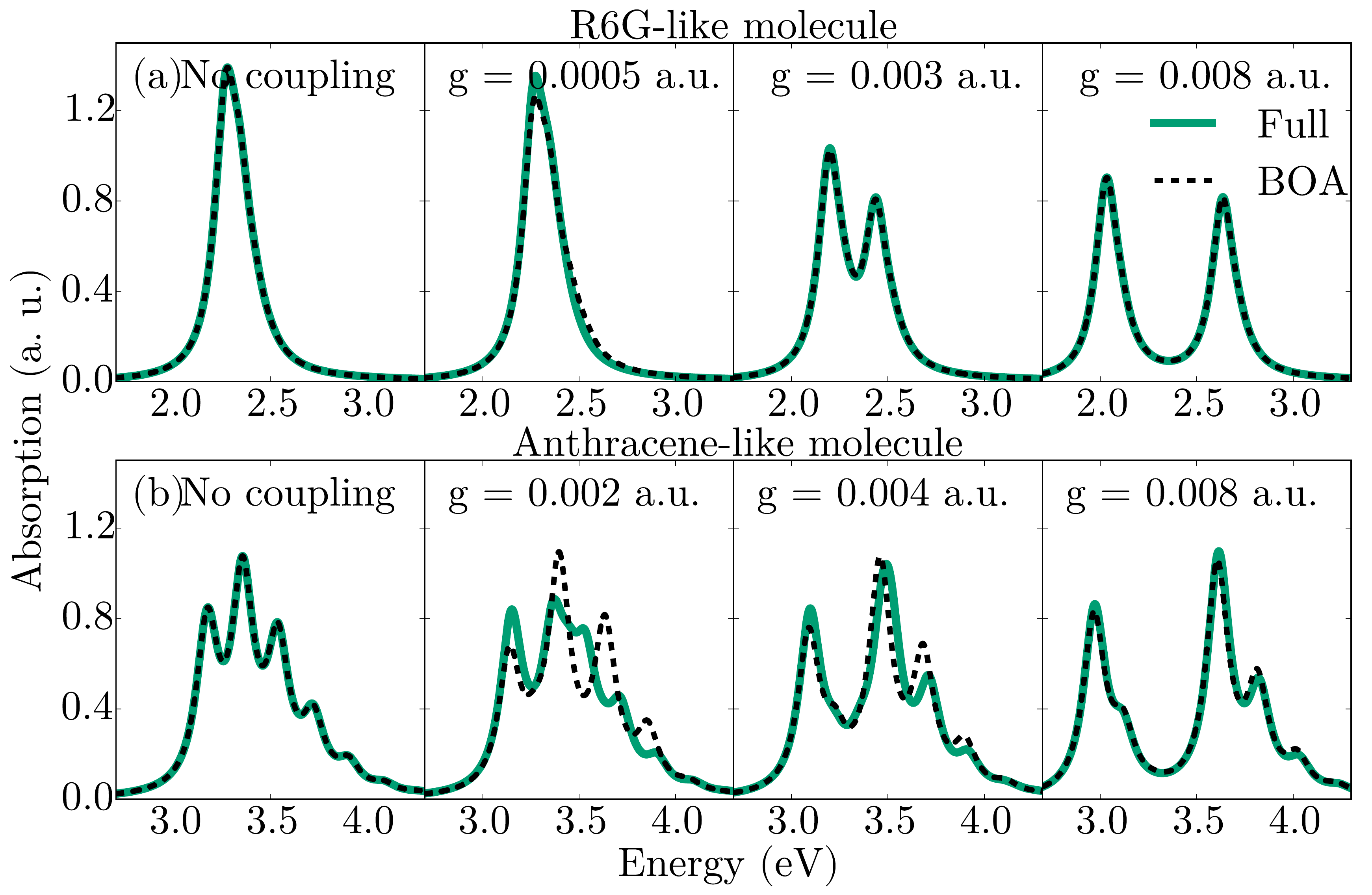}
\caption{Absorption cross sections of a single molecule, calculated using the full Hamiltonian without approximation (solid green lines) and within the BOA (dashed black lines). Results are shown for the (a) R6G-like and (b) anthracene-like model molecules, for several values of the coupling strength $g$.}
\label{fig:absorption_fullbo_AB}
\end{figure}

In \autoref{fig:absorption_fullbo_AB}, we compare the absorption cross sections under strong coupling as obtained from a full calculation without approximations to those obtained within the BOA, for a range of coupling strengths $g$ to the EM mode. Even for relatively small $g$, the BOA is found to agree almost perfectly with the full results for the R6G-like molecule with small vibrational spacing (\autoref{fig:absorption_fullbo_AB}a). However, for the anthracene-like molecule with a high-frequency vibrational mode and large offset $\DR$, the BOA only agrees with the full resul for relatively large values of $g$, where the Rabi splitting $\Omega_R$ (as defined by the energy difference between the two ``polariton'' peaks in the absorption spectrum) is appreciably larger than the vibrational frequency $\wvib\approx180$~meV (\autoref{fig:absorption_fullbo_AB}b).

This qualitative observation can be quantified by comparing the correction terms $\Cn^{kk'}$ in \autoref{eq:BOA_coupled} with the energy difference between the anticrossing PES at the point of closest approach. In \appref{app:non-BO}, we present a model that achieves this \emph{without} any explicit knowledge of the electronic wave functions. It relies on the observation that close to the anticrossing, the coupled (polariton) states switch character between the two uncoupled states, while the ``intrinsic'' $R$-dependence of the uncoupled electronic states can be neglected. The correction terms $\Cn^{kk'}$ can then be obtained just from the knowledge of $E_g(R)$, $E_e(R)$, and $\mu_{eg}(R)$, where $\mu_{eg}(R)$ is the electronic transition dipole between the ground and excited state. By approximating $E_g(R)$ and $E_e(R)$ as harmonic oscillators, the correction terms can be analytically evaluated and are found to be negligible under the condition that $\DR \wvib^2 / \Omega_R^2$ is small compared to the nuclear momentum of the relevant eigenstates. This demonstrates that the model molecules present two opposite cases for the applicability of the BO approximation: our R6G-like molecule has a relatively small vibrational spacing $\wvib\approx70$~meV and small electron-phonon coupling, $\DR\approx0.018$~a.u., while our anthracene-like model molecule has a large vibrational spacing $\wvib\approx180$~meV and large electron-phonon coupling, $\DR\approx0.092$~a.u.. We note that in many experiments involving organic molecules, $\Omega_R\gtrsim500$~meV~\cite{Schwartz2011,Kena-Cohen2013} is significantly larger than typical vibrational frequencies $\wvib\lesssim200$~meV~\cite{Shimanouchi1972}. This shows that the intuitive picture of nuclear dynamics unfolding on uncoupled Born-Oppenheimer potential energy surfaces can often be applied to understand the modification of molecular chemistry induced by strong coupling. Additionally, even when the BOA breaks down, the model presented in \appref{app:non-BO} can be used to obtain the non-BO coupling terms without requiring knowledge of the electronic wave functions. The only necessary input are the uncoupled PES and the associated transition dipole moments. Even for relatively large molecules, these can be obtained using the standard methods of quantum chemistry or density functional theory.

\section{Two molecules}\label{sec:two_molecules}
In the previous section, we showed that on the single-molecule level, the BOA is valid as long as the Rabi splitting $\Omega_R$ is large enough. However, current experiments are performed with large numbers of molecules, where coherent superpositions of electronic excitations (bright ``Dicke states''~\cite{Dicke1954}) couple strongly to the photonic mode(s), while other superpositions give uncoupled or ``dark'' modes. It is thus important to consider if and how our conclusions have to be modified when more than a single molecule is involved in strong coupling.

\begin{figure*}
\includegraphics[width=\linewidth]{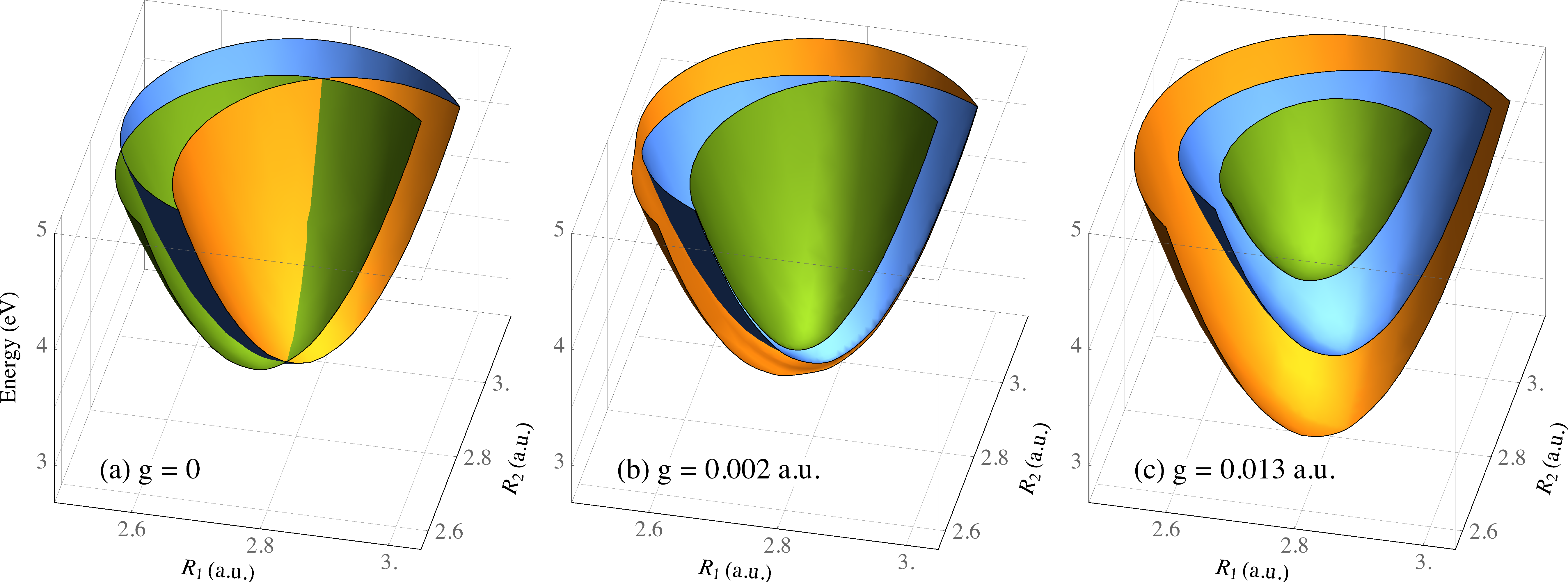}
\caption{(a) Uncoupled potential energy surfaces of two anthracene-like molecules in the singly excited subspace: $E_{eg0}(R_1,R_2)$ (orange), $E_{eg0}(R_1,R_2)$ (blue), and $E_{gg1}(R_1,R_2)$ (green). (b) Coupled PES for $g = 0.002$~a.u.\ and (c) $g = 0.013$~a.u., corresponding to the lower polariton (orange), dark state (blue), and upper polariton (green). For clarity, only parts where $R_1<R_2$ are shown (note that the system is symmetric under the exchange $R_1\leftrightarrow R_2$).}
\label{fig:two_surfaces}
\end{figure*}

For later reference, we give a quick overview of the theory when using an ensemble of two-level emitters coupled to a photonic mode, i.e., the many-particle Jaynes-Cummings (JC) model~\cite{Jaynes1963}, also known as the the Tavis-Cummings model~\cite{Tavis1967}. Its Hamiltonian within the rotating-wave approximation is
\begin{equation}\label{eq:JC_hamiltonian}
  \hat{H}_{JC}=\omega_c \hat{a}^{\dagger}\hat{a} + \sum_i \omega_i \hat{c}^{\dagger}_i \hat{c}_i + \sum_i g_i (\hat{a}\hat{c}^{\dagger}_i + \hat{a}^{\dagger}\hat{c}_i),
\end{equation}
where $\omega_i$ is the energy of emitter $i$ with destruction (creation) operator $\hat{c}_i$ ($\hat{c}_i^\dagger$), and the $g_i$ describe the emitter-photon couplings. For identical emitters ($\omega_i=\omega_m$, $g_i=g$), the resulting eigenstates in the single-excitation subspace are given by two polaritons $\ket{\pm} = \frac{1}{\sqrt{2}}(\hat{a}^{\dagger}\ket{0} \pm \ket{B})$, symmetric and antisymmetric combinations of the photonic mode with the emitter \emph{bright state} $\ket{B} = \frac{1}{\sqrt{N}} \sum_i \hat{c}_i ^{\dagger}\ket{0}$. At zero detuning ($\omega_c=\omega_m$), the polariton energies are given by $\omega_\pm = \omega_m \pm \Omega_R/2$, where $\Omega_R=2g\sqrt{N}$ is the collective Rabi splitting. The $N-1$ superpositions of emitter states orthogonal to $\ket{B}$ are \emph{dark states} that are not coupled to the photonic mode, with energies identical to the uncoupled emitters, $\omega_{DS}=\omega_m$. Note that in configurations with many photonic modes (e.g., planar cavities), more than one emitter state is coupled to the photonic mode (typically at low in-plane momentum), but there remain many uncoupled (dark) modes at higher in-plane momentum~\cite{Michetti2015,Agranovich2011}. There is an ongoing discussion in the literature on whether the ``dark'' modes are affected by strong coupling as well, or whether they should be thought of as completely unmodified emitter states. We will show below that when taking the internal structure of the emitters (molecules) into account, even the ``dark'' modes are affected by strong coupling and the nuclear dynamics of separate molecules become correlated.

We now treat the case of two model molecules, which can still be solved exactly within our approach, but which displays many of the effects of many-molecule strong coupling. As in the JC model, we assume that the two molecules both couple to the same photonic mode, but do not directly interact with each other, giving
\begin{equation}
\Hmc^{2m} = \omega_c \hat{a}^{\dagger} \hat{a} + \sum_{j=1,2} \left(\Hm^{(j)} + g \hat{\mu}^{(j)} (\hat{a}^{\dagger}+\hat{a}) \right),
\label{eq:twomol_hamiltonian}
\end{equation}
where the superscripts $j$ indicate the molecule on which the operator acts. Directly diagonalizing this Hamiltonian in the ``raw'' basis $\{x_1,R_1,x_2,R_2,n\}$ is already a formidable computational task for typical grid sizes. We thus calculate the full solution by first diagonalizing the single-molecule Hamiltonian, $\Hm=\sum_k E_k\ket{k}\bra{k}$, and including only a relevant subset of eigenstates $\{k\}$ for each molecule in the total basis $\{k_1,k_2,n\}$. The number of necessary eigenstates to obtain completely converged results is quite small ($\approx30$ per molecule). However, this approach only provides limited insight into the dynamics of the strongly coupled system, especially regarding nuclear motion.

We thus again apply the Born-Oppenheimer approximation by separating the nuclear kinetic energy terms and diagonalizing the remaining Hamiltonian parametrically as a function of $R_1$ and $R_2$. Similar to above, instead of working in the $\{x_1,x_2,n\}$ basis for each combination $(R_1,R_2)$, we prediagonalize the single-molecule electronic Hamiltonian $\He(x;R) = \sum_k E_k(R) \ket{k(R)}\bra{k(R)}$, where (for the cases discussed here) the sum only has to include the ground and first excited states to achieve convergence, $k\in\{g,e\}$. If we additionally allow at most one photon in the system, $n\in\{0,1\}$, we obtain an $8\times8$ Hamiltonian for each combination of nuclear coordinates $R_1$, $R_2$.

The electronic Hamiltonian consists of all possible combinations of electronic states $E_g$, $E_e$ of the two molecules with $0$ or $1$ photons. A further simplification is achieved by taking into account that the Hamiltonian conserves parity $\Pi = (-1)^{\pi_1+\pi_2+n}$, with $\pi_j$ the parity of the state of molecule $j$ (gerade or ungerade). For large coupling $g$, this separation by parity avoids some accidental degeneracies between uncoupled PES and thus improves the BOA. We now obtain two independent $4\times4$ Hamiltonians,
\begin{subequations} \label{eq:twomol_matrices}
\begin{align}
\hat{H}_{\mathrm{even}} (R_1,R_2) &= \begin{pmatrix}
E_{gg0} & gd^{(1)} & gd^{(2)} & 0 \\[0.1cm]
gd^{(1)} & E_{eg1} & 0 & gd^{(1)} \\[0.1cm]
gd^{(2)} & 0 & E_{ge1} & gd^{(2)} \\[0.1cm]
0 & gd^{(1)} & gd^{(2)} & E_{ee0}
\end{pmatrix}, \label{eq:twomol_even}\\[0.2cm]
\hat{H}_{\mathrm{odd}} (R_1,R_2) &= \begin{pmatrix}
E_{gg1} & gd^{(1)} & gd^{(2)} & 0 \\[0.1cm]
gd^{(1)} & E_{eg0} & 0 & gd^{(1)} \\[0.1cm]
gd^{(2)} & 0 & E_{ge0} & gd^{(2)} \\[0.1cm]
0 & gd^{(1)} & gd^{(2)} & E_{ee1}
\end{pmatrix},
\label{eq:twomol_odd}
\end{align}
\end{subequations}
where the uncoupled PES are represented by the compact notation $E_{abn} = E_a(R_1)+E_b(R_2)+n\omega_c$, and the single-molecule dipole transition moment between the ground and first excited state is denoted by $d^{(j)}=\bra{\phi_g(R_j)}\hat\mu\ket{\phi_e(R_j)}$. Diagonalizing these Hamiltonians for each $(R_1,R_2)$ results in a set of strongly coupled two-dimensional PES. In \autoref{fig:two_surfaces}, we show the three surfaces in the single-excitation subspace, corresponding to the three lowest states of \autoref{eq:twomol_odd}. For zero molecule-photon coupling ($g=0$, \autoref{fig:two_surfaces}a), there are now a number of one-dimensional seams where the three PES cross. When the molecule-photon coupling is turned on, these crossings again turn into avoided crossings, as shown in panels (b) and (c) for two different coupling strengths $g$.
Following the conventions used in the Jaynes-Cummings model, we label the three coupled PES in order of energy as the ``lower polariton PES'', the ``dark-state PES'', and the ``upper polariton PES''.

\begin{figure}
\includegraphics[width=\linewidth]{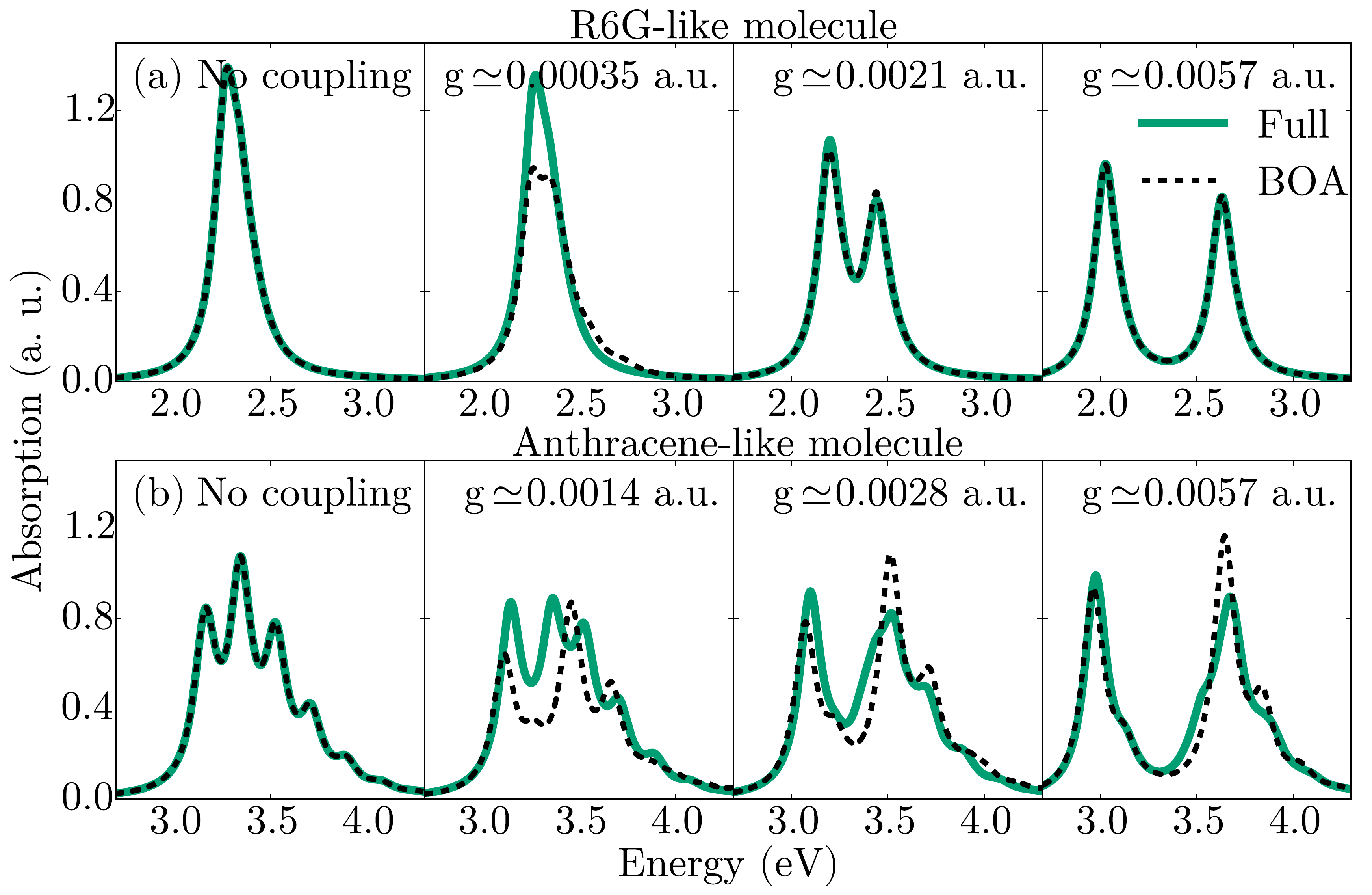}
\caption{Absorption cross section of two molecules driven coherently, calculated using the full Hamiltonian without approximation (solid green lines) and within the BOA (dashed black lines). Results are shown for the (a) R6G-like and (b) anthracene-like model molecules, for several values of the coupling strength $g$. The values of $g$ are scaled by $1/\sqrt{2}$ with respect to the single-molecule case (\autoref{fig:absorption_fullbo_AB}) in order to obtain the same total Rabi frequency $\Omega_R$.}
\label{fig:absorption_fullbo_2mol_AB}
\end{figure}

We first address the applicability of the BOA, which breaks down when two PES are close in energy, for the case of two molecules. Within the single-excitation subspace (which determines the linear properties of the system, such as absorption), there are now a range of (avoided) crossings. They occur when i) all three surfaces approach each other $E_{gg1} \approx E_{ge0} \approx E_{eg0}$, ii) the photonically excited PES is close to only one of the electronically excited PES, $E_{gg1} \approx E_{ge0}$ or $E_{gg1} \approx E_{eg0}$, or iii) only the two electronically excited states cross, $E_{ge0}\approx E_{eg0}$. Case i) corresponds to the JC model at zero detuning, giving the two polaritonic PES at energy shifts of $\pm \Omega_R/2$, and an additional dark state that is unshifted from the bare-molecule case. The BOA in this region is thus valid for similar conditions as in the single-molecule case, although the PES separation is reduced by half due to the additional dark-state surface. Case ii) corresponds exactly to the single-molecule case, with the second molecule acting as a ``spectator'' that only induces additional energy shifts. The BOA should thus again be valid for similar conditions as with a single-molecule, albeit with the coupling reduced by $1/\sqrt{2}$ for a fixed total Rabi splitting. Finally, case iii) presents the biggest challenge, as the two electronically excited PES, $E_{eg0}$ and $E_{ge0}$, are not directly coupled, but only split indirectly through coupling to the photonically excited surface $E_{gg1}$. The splitting between the two surfaces is thus small for large detuning, $\Delta E\approx (gd)^2/4(E_{gg1}-E_{eg0})$. This is clearly observed in \autoref{fig:two_surfaces}b along the line $R_1=R_2$, where the dark state PES almost touches the upper polariton PES for small $R$s and the lower polariton PES for large $R$s.

The discussion above implies that for almost any coupling strength, there will be regions in the nuclear configuration space $R_1,R_2$ where the BOA breaks down. However, not all parts of the PES are visited by the nuclei during a given physical process. To explicitly check the BOA in the subspace relevant for polaritonic physics, in \autoref{fig:absorption_fullbo_2mol_AB} we thus again compare the absorption with that obtained by a full diagonalization of the Hamiltonian \autoref{eq:twomol_hamiltonian}. Compared to the single-molecule case, many more molecular levels are present in the system, leading to small changes in the absorption spectra compared to the single-molecule case. In order to properly compare the results, we take into account the $\sqrt{N}$ scaling of the total Rabi frequency and reduce the coupling strengths by $\sqrt{2}$ to produce the same total splitting.
The BOA is shown to again become valid for large enough coupling, but the minimum coupling required is increased compared to that for a single molecule. In the common case of slow nuclear motion, as for our R6G-like model in \autoref{fig:absorption_fullbo_2mol_AB}a, the BOA already is valid for relatively small Rabi splitting of $\Omega_R\approx250$~eV. However, in the anthracene-like case of very fast vibrational motion, \autoref{fig:absorption_fullbo_2mol_AB}b, the BOA still does not give perfect agreement with the full model for $g=0.0057$~a.u.\ ($\Omega_R\approx600$~meV), and agreement is only reached at roughly twice that value.

Having established the validity of the BOA for many relevant cases and Rabi splittings comparable to experimental values, we now discuss the implications of collective strong coupling for the internal molecular (nuclear) dynamics. Note that this question can by design not be addressed within the JC model, where emitters are two-level systems without any internal degrees of freedom. In contrast, the BOA provides a straightforward approach to this problem. Any two-dimensional PES can be decomposed into a sum of independent single-molecule potentials, plus a remainder that describes the coupling between the nuclear motion of the molecules,
\begin{equation}
  E(R_1,R_2) = E_1(R_1) + E_2(R_2) + E_{12}(R_1,R_2) .
  \label{eq:coupled_PES_decomposition}
\end{equation}

\begin{figure}
\includegraphics[width=\linewidth]{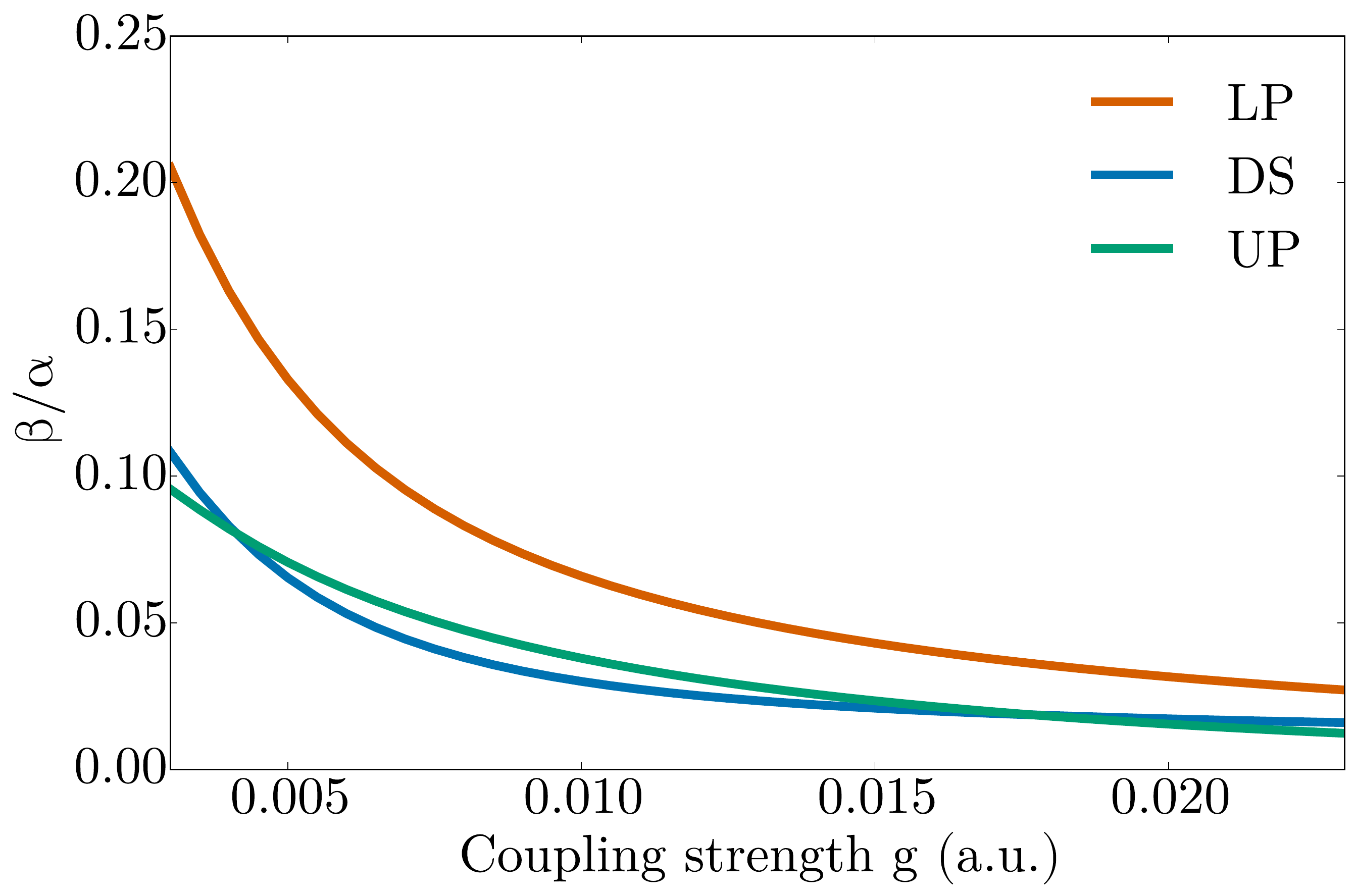}
\caption{Coupling between nuclear motion in different molecules for the lower (LP) and upper polariton (UP) and dark-state (DS) PES. Results are shown as the ratio $\beta/\alpha$ between the prefactors of the offdiagonal $R_1 R_2$ and diagonal $R_i^2$ terms in \autoref{eq:coupled_PES_decomposition_taylor}, for the R6G-like model molecule.}
\label{fig:coupled_PES_correlation}
\end{figure}

The nuclear motion of two molecules is independent if and only if the coupled part $E_{12}(R_1,R_2)$ is identically zero. In order to quantify this coupling, we expand each of the coupled PES in the single-excitation subspace around its minimum $(R_1^0,R_2^0)$, giving
\begin{equation}
  E(R_1,R_2) \approx E_0 + \alpha\,\dR_1^2 + \alpha\,\dR_2^2 + \beta\, \dR_1 \dR_2,
  \label{eq:coupled_PES_decomposition_taylor}
\end{equation}
with $E_0=E(R_1^0,R_2^0)$ and $\dR_i=R_i-R_i^0$. Note that due to symmetry under the exchange $R_1\leftrightarrow R_2$, the prefactor $\alpha$ is the same for $\dR_1^2$ and $\dR_2^2$. As can be seen in \autoref{fig:coupled_PES_correlation}, both the polariton and even the dark state PES show significant coupling of the nuclear degrees of freedom, with values of $\beta/\alpha$ on the order of a few percent for values of $g\lesssim0.01$~a.u.\ giving realistic Rabi splittings of $\lesssim1$~eV (see~\autoref{fig:absorption_fullbo_2mol_AB}). Interestingly, the coupling is much larger for the lower polariton state than for either the upper polariton or the dark state, and decreases with increasing $g$ for all three PES. We therefore conclude that even dark states that have negligible mixing with photonic modes are affected by strong coupling, in the sense that the nuclear degrees of freedom of separate molecules behave like coupled harmonic oscillators, and their motion becomes correlated. This implies that, e.g., local excitation of nuclear motion within one molecule could affect the nuclear motion in another, spatially separated molecule, even when no photon is ever present in the system. Note that these results apply within the singly-excited subspace, i.e., the coupled nuclear motion is only observed when electronic excitation is present, not in the ground state. In the next section, we discuss which modifications of the ground state properties could be observed in the ultrastrong coupling regime.

\section{Ultrastrong coupling and ground state modifications}\label{sec:ultrastrong_coupling}
Up to now, we focused on the molecular properties in the singly excited subspace, which are probed in linear response measurements such as absorption and transmission, and where the effect of strong coupling is immediately apparent. However, when the Rabi frequency, i.e., the energy exchange rate between the molecules and the photonic mode, becomes significant compared to the frequencies of these two modes, the so-called \emph{ultra-strong} coupling regime is entered~\cite{DeLiberato2007,Carusotto2013}. In this regime, the rotating wave approximation for the emitter-light interaction (under which the number of excitations is conserved) becomes invalid. In our approach, the rotating wave approximation is not used, and we can thus naturally explore the ultrastrong coupling regime. One of its most intriguing predictions is that even the ground-state properties of the system should be significantly modified. For example, the ground state is shifted in energy and acquires a photonic component, such that a number of virtual photons are present in the system even without any external excitations. This raises the question of how the internal degrees of freedom of organic molecules are affected when this regime is entered.

The BOA is well-suited to explore this regime. In contrast to the singly-excited subspace, where narrow avoided crossings can affect its validity, the ground-state PES is energetically well-separated from all other PES. This remains true even under ultrastrong coupling, and consequently the BOA is valid for all coupling strengths. The ground state potential energy surface $E_g(R)$ is coupled to the doubly excited surface $E_e(R)+\omega_c$ (cf.~\autoref{eq:twomol_even}), with the strongly coupled ground state PES given to lowest order by $E_g^{SC}(R) \approx E_g(R) - \frac{g^2\mu_{eg}^2(R)}{E_e(R)+\omega_c-E_g(R)} + O(g^4)$. Ground state properties such as the bond length are determined by the shape of the PES. The largest modification can thus be expected when the $R$-dependence of the ground and excited PES is as different as possible. This occurs for large electron-phonon coupling, i.e., a large value of $\DR$, such as in our anthracene-like molecule. For a coupling strength of $g=0.016$~a.u., corresponding to a Rabi splitting of $\Omega_R\approx1.2$~eV in absorption, we obtain a shift in the ground state bond length of $\DR_0\approx0.84$~mÅ$=84$~fm. While small, such a change in bond length could be detectable using X-ray absorption fine structure spectroscopy or X-ray crystallography, which can obtain few- or even sub-femtometer precision for measuring bond length shifts~\cite{Konig2012,Kozina2014}.

However, the previous paragraph only applies for a single molecule under strong coupling. Repeating the calculation using two molecules and taking into account the reduced single-molecule coupling strength (for fixed Rabi splitting, $\Omega_R\propto \sqrt{N}g$) reveals a reduction of the bond-length change by a factor of two, $\DR_0^{2mol}\approx0.42$~mÅ. This is confirmed by using a similar analytical model as presented in \appref{app:non-BO}, in which the bare-molecule potential energy surfaces are approximated as harmonic oscillators. Due to the simple analytical structure, perturbation theory can be applied to obtain results for arbitrary numbers of molecules, and the change in ground-state bond length is found to be proportional to the squared \emph{single-molecule} coupling $g^2$, not to the squared collective Rabi frequency $\Omega_R^2$. We note that in contrast, the ground-state energy shift is indeed determined by the collective coupling strength, $\Delta E_0\propto\Omega_R^2$. In realistic experimental configurations involving large numbers of molecules, the change in ground state bond length is thus expected to be minuscule and extremely challenging to measure. This demonstrates that the influence of strong coupling on any specific observable is not immediately obvious, and has to be checked case by case. For some properties, the molecules will behave as if they feel the full collective coupling $\Omega_R$, while for others, they will only show the change induced by the single-molecule coupling $g$.

We thus check another observable, and ask whether the ground state will show correlated nuclear motion between distant molecules, as observed in the dark state surface. This can again be quantified using the expansion of the PES in \autoref{eq:coupled_PES_decomposition_taylor}. Doing so reveals a very small coupling parameter $\beta$ that to lowest order is proportional to $g^4/\omega_c^5$ (close to zero detuning, $\omega_c\sim E_e(R)-E_g(R)$). This corresponds to an even higher-order correction than the already small energy or bond-length shifts. Furthermore, like the bond-length shift, it depends on the single-molecule coupling instead of the collective coupling strength. We can thus conclude that in contrast to the excited states, the ground-state nuclear motion of the molecules does not become correlated even in the ultrastrong coupling limit.

\section{Conclusions \& Outlook}
We have presented a simple model that offers a new perspective on strong coupling with organic molecules. We have shown under which conditions the molecular properties under strong coupling can be understood by the modification of the potential energy surfaces determining nuclear dynamics under the Born-Oppenheimer approximation. In particular, we found that in many cases of experimental interest where the Rabi splitting is large, the BOA is applicable and provides an intuitive picture of the strongly coupled dynamics. However, we have also shown that for molecules with fast vibrational modes and large phonon-exciton couplings, the BOA can break down and a more involved picture is necessary. We furthermore demonstrated that the non-BO coupling terms between PES in this case are dominantly due to the change of character between light and matter excitations which can be obtained from simple few-level models without requiring knowledge of the electronic wavefunctions.

In addition, we have shown that under collective strong coupling involving more than one molecule, the nuclear dynamics of the molecules in electronic ``dark states'' that are only weakly coupled to the photonic mode are nonetheless affected by the formation of strong coupling. In particular, we find that the dark state PES describes coupling between the nuclear degrees of freedom of the different molecules.

Finally, we investigated the change of the ground state properties under ultrastrong coupling, where the Rabi splitting becomes a significant fraction of the transition energy. Using our numerical calculations and a simple analytical model, we showed that while the ground-state energy shift is affected by the collective Rabi frequency (which is enhanced by $\sqrt{N}$ for $N$ molecules), other properties such as the ground-state bond length depend on the single-molecule coupling strength and are not significantly affected for experimentally realistic parameters.

Our results also lay the groundwork for a further in-depth exploration of the modification of molecular properties under strong coupling. In particular, they provide a simple picture that can be used to understand the modification of chemical reactions, e.g., by lowering a potential barrier along a reaction coordinate. There are also a number of obvious extensions of the simple model presented here that will be explored in the future. These include more realistic models of organic molecules using more degrees of freedom (for example, employing the PES obtained using quantum chemistry packages), and the inclusion of incoherent processes such as decay and decoherence. We note that there has been some recent progress on combining QED with density functional theory~\cite{Tokatly2013,Ruggenthaler2014}, which could provide complementary information to the model presented here.

\begin{acknowledgments}
This work has been funded by the European Research Council (ERC-2011-AdG proposal No. 290981), by the European Union Seventh Framework Programme under grant agreement FP7-PEOPLE-2013-CIG-618229, and the Spanish MINECO under contract MAT2011-28581-C02-01.
\end{acknowledgments}

\appendix
\section{Model for non-Born-Oppenheimer corrections}\label{app:non-BO}
In this appendix, we derive an analytical model for the non-Born-Oppenheimer corrections $\Cn^{kk'}$ under molecular strong coupling, for a single molecule. We treat the two PES in the single-excitation subspace, $E_g(R)+\omega_c$ and $E_e(R)$, coupled by the term $g\mu_{eg}(R)$. This leads to a $2\times2$ BO Hamiltonian of the form
\begin{equation}
\hat{H}(R) = \begin{pmatrix}
E_g(R)+\omega_c & g\mu_{eg}(R) \\
g\mu_{eg}(R) & E_e(R)
\end{pmatrix},
\end{equation}
which can be easily diagonalized to obtain polariton eigenstates $\ket{+} = \cos\theta \ket{g1} + \sin\theta \ket{e0}$ and $\ket{-} = \sin\theta \ket{g1} - \cos\theta \ket{e0}$, where $\ket{an}$ is short for $\ket{\phi_a(x;R),n}$, and
\begin{equation}
\tan2\theta = \frac{2h(R)} {\delta E(R)}\,,
\end{equation}
where we defined $\delta E(R) = E_g(R)+\omega_c-E_e(R)$ and $h(R)=g\mu_{eg}(R)$. Using $E_{\mathrm{avg}}(R) = \frac{E_g(R)+\omega_c+E_e(R)}{2}$, the eigenenergies are given by 
\begin{equation}
  E_\pm(R) = E_{\mathrm{avg}}(R) \pm \frac12 \sqrt{4h^2(R) + \delta E(R)^2} \,.
\end{equation}

We can now evaluate the non-Born-Oppenheimer coupling terms $\Cn^{kk'} = \bra{k} \Tn \ket{k'} + \bra{k} \frac{\hat{P}}{2M} \ket{k'} \hat{P}$ within this model, using a series of approximations to obtain simple analytical results. First, we linearize $\delta E(R) \approx a_0 (R-R_c)$ around the point of intersection between the two PES, where $E_g(R_c)+\omega_c = E_e(R_c)$. Second, in the spirit of the Franck-Condon approximation, we assume that the dipole coupling is constant over the range of relevant $R$-values, and set $h(R)=h_0$. Following the same idea, we additionally assume that the electronic wave functions do not change significantly with $R$, i.e., $\frac{\partial}{\partial R} \ket{\phi_a(x;R)}\approx 0$. This implies that the change in the polaritonic eigenfunctions $\ket{\pm}$ close to the avoided crossing at $R_c$ is mostly due to the switchover between the two uncoupled surfaces, i.e., the change in $\theta(R)$, not because of an intrinsic change of electronic state with $R$. With these approximations, the correction terms become
\begin{subequations}\label{eq:non-BO-coupling-model}
\begin{align}
\bra{-} \hat{P} \ket{+} &= \frac{-i a_0 h_0}{4 h_0^2 + a_0^2 (R-R_c)^2} ,\\
\bra{-} \hat{P}^2 \ket{+} &= \frac{2a_0^3 h_0 (R-R_c)}{(4 h_0^2 + a_0^2 (R-R_c)^2)^2},\\
\bra{\pm} \hat{P}^2 \ket{\pm} &= \frac{a_0^2 h_0^2}{(4 h_0^2 + a_0^2 (R - R_c)^2)^2},
\end{align}
\end{subequations}
with the diagonal terms $\bra{\pm}\hat{P}\ket{\pm}$ identically zero. Note that diagonal terms only correspond to energy shifts and do not induce additional transitions~\cite{Tully2000}. The non-Born-Oppenheimer coupling between the polariton surfaces has a Lorentzian shape around the avoided crossing, and as expected only becomes non-negligible close to it.
\begin{figure}
\includegraphics[width=\linewidth]{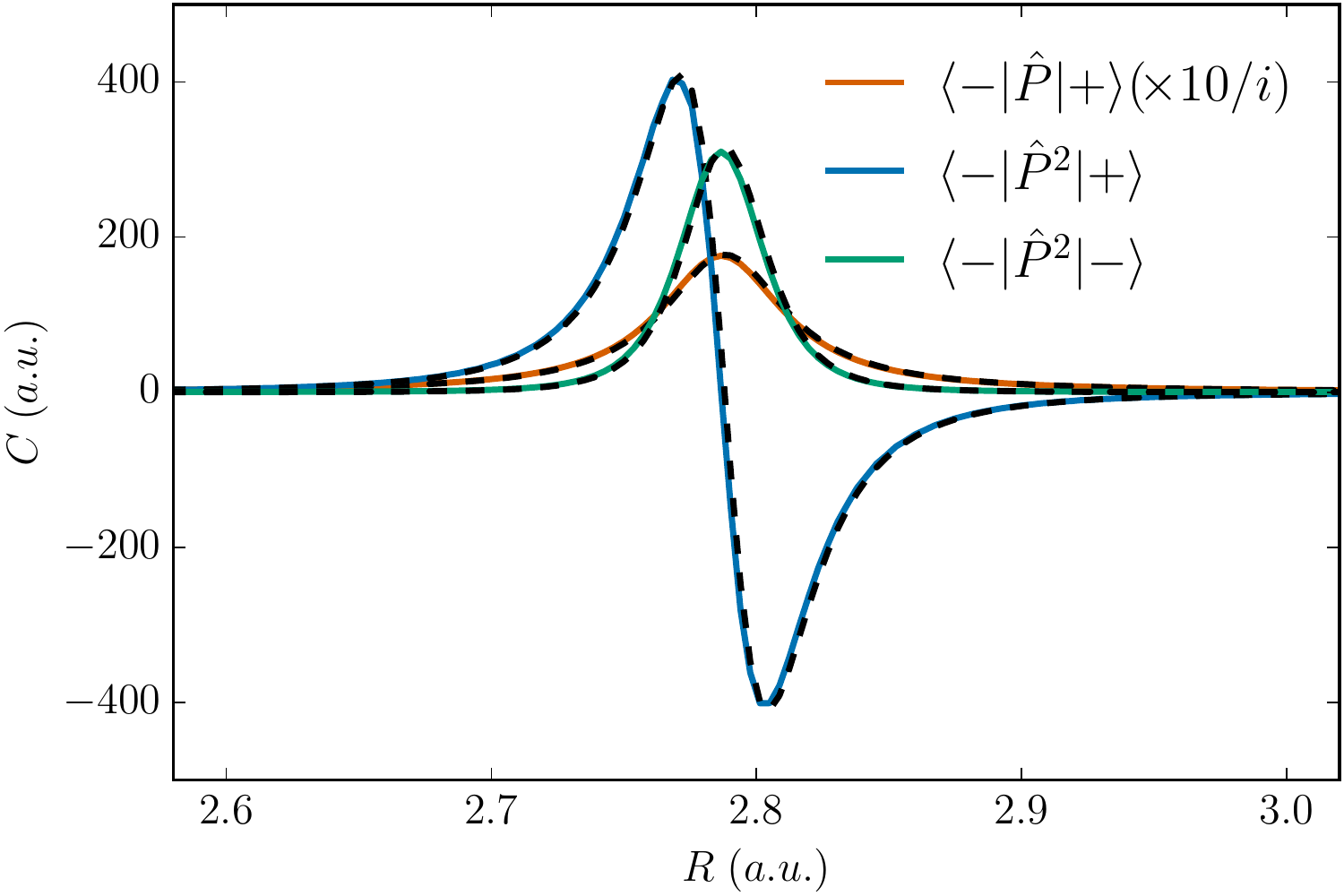}
\caption{Non-Born-Oppenheimer correction terms coupling the ``lower polariton'' and ``upper polariton'' PES for a single anthracene-like model molecule for a coupling strength of $g=0.002$~a.u.. Solid colored lines: results from a full numerical calculation. Dashed black lines: results from the model \autoref{eq:non-BO-coupling-model}. Note that while all results are given in atomic units, the units of the $\hat{P}$ and $\hat{P}^2$ terms are not identical, and thus not directly comparable.}
\label{fig:nonBO_corrections}
\end{figure}
As shown in \autoref{fig:nonBO_corrections}, the non-Born-Oppenheimer corrections obtained from this simple model agree almost perfectly with those obtained from the full numerical calculation for our anthracene-like model molecule. The only molecule-specific information entering the model are the PES of the uncoupled molecule and the dipole moment between the coupled surfaces. Specifically, the \emph{electronic} wave functions are never used here, and their derivative as a function of the nuclear coordinates is not required. This implies that this model could be used to obtain accurate non-BO corrections that describe the transitions between potential surfaces even when the full electronic wave functions of a molecule are not available (e.g., in DFT calculations). The dynamics of the molecule could thus be fully recovered within a potential energy surface picture even when the BOA per se is not applicable.

We now exploit this model to derive a condition for when the BOA becomes applicable, i.e., when the non-BO terms become negligible. We approximate the bare molecular potential energy surfaces as two harmonic oscillators with the same vibrational frequency $\wvib$, but with an offset in energy and equilibrium position,
\begin{align}
E_g(R) &\approx \frac{M\wvib^2}{2} R^2 ,\\
E_e(R) &\approx \frac{M\wvib^2}{2} (R-\DR)^2 + \Delta E,
\end{align}
where without loss of generality, we have chosen the origin in $R$ and $E$ at the minimum of $E_g(R)$. Note that this model exactly results from the common approximation of linear coupling between a single electronic excitation and a bosonic vibrational mode~\cite{Leggett1987,Coalson1994}. Within this model, $\delta E(R)=E_g(R)+\omega_c-E_e(R) = a_0 (R-R_c)$ is already exactly linear, i.e., the linearization of the energy difference performed above is not an approximation. The constants are given by $a_0=M\wvib^2 \DR$ and $R_c=\frac{\DR}{2} + \frac{\Delta E-\omega_c}{a_0}$. The maximum value of $|\bra{+} \frac{\hat{P}}{2M} \ket{-}|$, reached at $R=R_c$, is given by $\DR \wvib^2/(8h_0)$. Comparing this with the energy splitting at that point, $E_+(R_c)-E_-(R_c)=2h_0$, gives the condition that $\DR \wvib^2/(16 h_0^2)$ must be small compared to the momentum of the respective nuclear wavefunction (due to the additional $\hat{P}$ operating on the nuclear wave function). The off-diagonal terms from $\bra{-} \hat{P}^2 \ket{+}$ reach a maximum value (again relative to the detuning) of $M \DR^2 \wvib^4/(25 \sqrt{5} h_0^3)$ at $R=R_c+h_0/(M \DR \wvib^2)$.

\bibliography{mendeley}

\begin{thebibliography}{49}%
\makeatletter
\providecommand \@ifxundefined [1]{%
 \@ifx{#1\undefined}
}%
\providecommand \@ifnum [1]{%
 \ifnum #1\expandafter \@firstoftwo
 \else \expandafter \@secondoftwo
 \fi
}%
\providecommand \@ifx [1]{%
 \ifx #1\expandafter \@firstoftwo
 \else \expandafter \@secondoftwo
 \fi
}%
\providecommand \natexlab [1]{#1}%
\providecommand \enquote  [1]{``#1''}%
\providecommand \bibnamefont  [1]{#1}%
\providecommand \bibfnamefont [1]{#1}%
\providecommand \citenamefont [1]{#1}%
\providecommand \href@noop [0]{\@secondoftwo}%
\providecommand \href [0]{\begingroup \@sanitize@url \@href}%
\providecommand \@href[1]{\@@startlink{#1}\@@href}%
\providecommand \@@href[1]{\endgroup#1\@@endlink}%
\providecommand \@sanitize@url [0]{\catcode `\\12\catcode `\$12\catcode
  `\&12\catcode `\#12\catcode `\^12\catcode `\_12\catcode `\%12\relax}%
\providecommand \@@startlink[1]{}%
\providecommand \@@endlink[0]{}%
\providecommand \url  [0]{\begingroup\@sanitize@url \@url }%
\providecommand \@url [1]{\endgroup\@href {#1}{\urlprefix }}%
\providecommand \urlprefix  [0]{URL }%
\providecommand \Eprint [0]{\href }%
\providecommand \doibase [0]{http://dx.doi.org/}%
\providecommand \selectlanguage [0]{\@gobble}%
\providecommand \bibinfo  [0]{\@secondoftwo}%
\providecommand \bibfield  [0]{\@secondoftwo}%
\providecommand \translation [1]{[#1]}%
\providecommand \BibitemOpen [0]{}%
\providecommand \bibitemStop [0]{}%
\providecommand \bibitemNoStop [0]{.\EOS\space}%
\providecommand \EOS [0]{\spacefactor3000\relax}%
\providecommand \BibitemShut  [1]{\csname bibitem#1\endcsname}%
\let\auto@bib@innerbib\@empty
%</preamble>
\bibitem [{\citenamefont {Thompson}\ \emph {et~al.}(1992)\citenamefont
  {Thompson}, \citenamefont {Rempe},\ and\ \citenamefont
  {Kimble}}]{Thompson1992}%
  \BibitemOpen
  \bibfield  {author} {\bibinfo {author} {\bibfnamefont {R.~J.}\ \bibnamefont
  {Thompson}}, \bibinfo {author} {\bibfnamefont {G.}~\bibnamefont {Rempe}}, \
  and\ \bibinfo {author} {\bibfnamefont {H.~J.}\ \bibnamefont {Kimble}},\
  }\bibfield  {title} {\enquote {\bibinfo {title} {{Observation of normal-mode
  splitting for an atom in an optical cavity}},}\ }\href {\doibase
  10.1103/PhysRevLett.68.1132} {\bibfield  {journal} {\bibinfo  {journal}
  {Phys. Rev. Lett.}\ }\textbf {\bibinfo {volume} {68}},\ \bibinfo {pages}
  {1132--1135} (\bibinfo {year} {1992})}\BibitemShut {NoStop}%
\bibitem [{\citenamefont {Weisbuch}\ \emph {et~al.}(1992)\citenamefont
  {Weisbuch}, \citenamefont {Nishioka}, \citenamefont {Ishikawa},\ and\
  \citenamefont {Arakawa}}]{Weisbuch1992}%
  \BibitemOpen
  \bibfield  {author} {\bibinfo {author} {\bibfnamefont {C.}~\bibnamefont
  {Weisbuch}}, \bibinfo {author} {\bibfnamefont {M.}~\bibnamefont {Nishioka}},
  \bibinfo {author} {\bibfnamefont {A.}~\bibnamefont {Ishikawa}}, \ and\
  \bibinfo {author} {\bibfnamefont {Y.}~\bibnamefont {Arakawa}},\ }\bibfield
  {title} {\enquote {\bibinfo {title} {{Observation of the coupled
  exciton-photon mode splitting in a semiconductor quantum microcavity}},}\
  }\href {\doibase 10.1103/PhysRevLett.69.3314} {\bibfield  {journal} {\bibinfo
   {journal} {Phys. Rev. Lett.}\ }\textbf {\bibinfo {volume} {69}},\ \bibinfo
  {pages} {3314--3317} (\bibinfo {year} {1992})}\BibitemShut {NoStop}%
\bibitem [{\citenamefont {Kasprzak}\ \emph {et~al.}(2006)\citenamefont
  {Kasprzak}, \citenamefont {Richard}, \citenamefont {Kundermann},
  \citenamefont {Baas}, \citenamefont {Jeambrun}, \citenamefont {Keeling},
  \citenamefont {Marchetti}, \citenamefont {Szymańska}, \citenamefont
  {Andr\'{e}}, \citenamefont {Staehli}, \citenamefont {Savona}, \citenamefont
  {Littlewood}, \citenamefont {Deveaud},\ and\ \citenamefont
  {Dang}}]{Kasprzak2006}%
  \BibitemOpen
  \bibfield  {author} {\bibinfo {author} {\bibfnamefont {J.}~\bibnamefont
  {Kasprzak}}, \bibinfo {author} {\bibfnamefont {M.}~\bibnamefont {Richard}},
  \bibinfo {author} {\bibfnamefont {S.}~\bibnamefont {Kundermann}}, \bibinfo
  {author} {\bibfnamefont {A.}~\bibnamefont {Baas}}, \bibinfo {author}
  {\bibfnamefont {P.}~\bibnamefont {Jeambrun}}, \bibinfo {author}
  {\bibfnamefont {J.~M.~J.}\ \bibnamefont {Keeling}}, \bibinfo {author}
  {\bibfnamefont {F.~M.}\ \bibnamefont {Marchetti}}, \bibinfo {author}
  {\bibfnamefont {M.~H.}\ \bibnamefont {Szymańska}}, \bibinfo {author}
  {\bibfnamefont {R.}~\bibnamefont {Andr\'{e}}}, \bibinfo {author}
  {\bibfnamefont {J.~L.}\ \bibnamefont {Staehli}}, \bibinfo {author}
  {\bibfnamefont {V.}~\bibnamefont {Savona}}, \bibinfo {author} {\bibfnamefont
  {P.~B.}\ \bibnamefont {Littlewood}}, \bibinfo {author} {\bibfnamefont
  {B.}~\bibnamefont {Deveaud}}, \ and\ \bibinfo {author} {\bibfnamefont
  {Le~Si}\ \bibnamefont {Dang}},\ }\bibfield  {title} {\enquote {\bibinfo
  {title} {{Bose-Einstein condensation of exciton polaritons}},}\ }\href
  {\doibase 10.1038/nature05131} {\bibfield  {journal} {\bibinfo  {journal}
  {Nature}\ }\textbf {\bibinfo {volume} {443}},\ \bibinfo {pages} {409--14}
  (\bibinfo {year} {2006})}\BibitemShut {NoStop}%
\bibitem [{\citenamefont {Balili}\ \emph {et~al.}(2007)\citenamefont {Balili},
  \citenamefont {Hartwell}, \citenamefont {Snoke}, \citenamefont {Pfeiffer},\
  and\ \citenamefont {West}}]{Balili2007}%
  \BibitemOpen
  \bibfield  {author} {\bibinfo {author} {\bibfnamefont {R.}~\bibnamefont
  {Balili}}, \bibinfo {author} {\bibfnamefont {V.}~\bibnamefont {Hartwell}},
  \bibinfo {author} {\bibfnamefont {D.}~\bibnamefont {Snoke}}, \bibinfo
  {author} {\bibfnamefont {L.}~\bibnamefont {Pfeiffer}}, \ and\ \bibinfo
  {author} {\bibfnamefont {K.}~\bibnamefont {West}},\ }\bibfield  {title}
  {\enquote {\bibinfo {title} {{Bose-Einstein condensation of microcavity
  polaritons in a trap}},}\ }\href {\doibase 10.1126/science.1140990}
  {\bibfield  {journal} {\bibinfo  {journal} {Science}\ }\textbf {\bibinfo
  {volume} {316}},\ \bibinfo {pages} {1007--10} (\bibinfo {year}
  {2007})}\BibitemShut {NoStop}%
\bibitem [{\citenamefont {K\'{e}na-Cohen}\ and\ \citenamefont
  {Forrest}(2010)}]{Kena-Cohen2010}%
  \BibitemOpen
  \bibfield  {author} {\bibinfo {author} {\bibfnamefont {S.}~\bibnamefont
  {K\'{e}na-Cohen}}\ and\ \bibinfo {author} {\bibfnamefont {S.~R.}\
  \bibnamefont {Forrest}},\ }\bibfield  {title} {\enquote {\bibinfo {title}
  {{Room-temperature polariton lasing in an organic single-crystal
  microcavity}},}\ }\href {\doibase 10.1038/nphoton.2010.86} {\bibfield
  {journal} {\bibinfo  {journal} {Nat. Phot.}\ }\textbf {\bibinfo {volume}
  {4}},\ \bibinfo {pages} {371--375} (\bibinfo {year} {2010})}\BibitemShut
  {NoStop}%
\bibitem [{\citenamefont {Plumhof}\ \emph {et~al.}(2014)\citenamefont
  {Plumhof}, \citenamefont {St\"{o}ferle}, \citenamefont {Mai}, \citenamefont
  {Scherf},\ and\ \citenamefont {Mahrt}}]{Plumhof2014}%
  \BibitemOpen
  \bibfield  {author} {\bibinfo {author} {\bibfnamefont {Johannes~D.}\
  \bibnamefont {Plumhof}}, \bibinfo {author} {\bibfnamefont {Thilo}\
  \bibnamefont {St\"{o}ferle}}, \bibinfo {author} {\bibfnamefont {Lijian}\
  \bibnamefont {Mai}}, \bibinfo {author} {\bibfnamefont {Ullrich}\ \bibnamefont
  {Scherf}}, \ and\ \bibinfo {author} {\bibfnamefont {Rainer~F.}\ \bibnamefont
  {Mahrt}},\ }\bibfield  {title} {\enquote {\bibinfo {title} {{Room-temperature
  Bose-Einstein condensation of cavity exciton-polaritons in a polymer.}}}\
  }\href {\doibase 10.1038/nmat3825} {\bibfield  {journal} {\bibinfo  {journal}
  {Nat. Mater.}\ }\textbf {\bibinfo {volume} {13}},\ \bibinfo {pages} {247--52}
  (\bibinfo {year} {2014})}\BibitemShut {NoStop}%
\bibitem [{\citenamefont {Daskalakis}\ \emph {et~al.}(2014)\citenamefont
  {Daskalakis}, \citenamefont {Maier}, \citenamefont {Murray},\ and\
  \citenamefont {K\'{e}na-Cohen}}]{Daskalakis2014}%
  \BibitemOpen
  \bibfield  {author} {\bibinfo {author} {\bibfnamefont {K.~S.}\ \bibnamefont
  {Daskalakis}}, \bibinfo {author} {\bibfnamefont {S.~A.}\ \bibnamefont
  {Maier}}, \bibinfo {author} {\bibfnamefont {R.}~\bibnamefont {Murray}}, \
  and\ \bibinfo {author} {\bibfnamefont {S.}~\bibnamefont {K\'{e}na-Cohen}},\
  }\bibfield  {title} {\enquote {\bibinfo {title} {{Nonlinear interactions in
  an organic polariton condensate.}}}\ }\href {\doibase 10.1038/nmat3874}
  {\bibfield  {journal} {\bibinfo  {journal} {Nat. Mater.}\ }\textbf {\bibinfo
  {volume} {13}},\ \bibinfo {pages} {271--8} (\bibinfo {year}
  {2014})}\BibitemShut {NoStop}%
\bibitem [{\citenamefont {Lidzey}\ \emph {et~al.}(1998)\citenamefont {Lidzey},
  \citenamefont {Bradley}, \citenamefont {Skolnick}, \citenamefont {Virgili},
  \citenamefont {Walker},\ and\ \citenamefont {Whittaker}}]{Lidzey1998}%
  \BibitemOpen
  \bibfield  {author} {\bibinfo {author} {\bibfnamefont {D.~G.}\ \bibnamefont
  {Lidzey}}, \bibinfo {author} {\bibfnamefont {D.~D.~C.}\ \bibnamefont
  {Bradley}}, \bibinfo {author} {\bibfnamefont {M.~S.}\ \bibnamefont
  {Skolnick}}, \bibinfo {author} {\bibfnamefont {T.}~\bibnamefont {Virgili}},
  \bibinfo {author} {\bibfnamefont {S.}~\bibnamefont {Walker}}, \ and\ \bibinfo
  {author} {\bibfnamefont {D.~M.}\ \bibnamefont {Whittaker}},\ }\bibfield
  {title} {\enquote {\bibinfo {title} {{Strong exciton-photon coupling in an
  organic semiconductor microcavity}},}\ }\href {\doibase 10.1038/25692}
  {\bibfield  {journal} {\bibinfo  {journal} {Nature}\ }\textbf {\bibinfo
  {volume} {395}},\ \bibinfo {pages} {53--55} (\bibinfo {year}
  {1998})}\BibitemShut {NoStop}%
\bibitem [{\citenamefont {Schwartz}\ \emph {et~al.}(2011)\citenamefont
  {Schwartz}, \citenamefont {Hutchison}, \citenamefont {Genet},\ and\
  \citenamefont {Ebbesen}}]{Schwartz2011}%
  \BibitemOpen
  \bibfield  {author} {\bibinfo {author} {\bibfnamefont {T.}~\bibnamefont
  {Schwartz}}, \bibinfo {author} {\bibfnamefont {J.~A.}\ \bibnamefont
  {Hutchison}}, \bibinfo {author} {\bibfnamefont {C.}~\bibnamefont {Genet}}, \
  and\ \bibinfo {author} {\bibfnamefont {T.~W.}\ \bibnamefont {Ebbesen}},\
  }\bibfield  {title} {\enquote {\bibinfo {title} {{Reversible Switching of
  Ultrastrong Light-Molecule Coupling}},}\ }\href {\doibase
  10.1103/PhysRevLett.106.196405} {\bibfield  {journal} {\bibinfo  {journal}
  {Phys. Rev. Lett.}\ }\textbf {\bibinfo {volume} {106}},\ \bibinfo {pages}
  {196405} (\bibinfo {year} {2011})}\BibitemShut {NoStop}%
\bibitem [{\citenamefont {K\'{e}na-Cohen}\ \emph {et~al.}(2013)\citenamefont
  {K\'{e}na-Cohen}, \citenamefont {Maier},\ and\ \citenamefont
  {Bradley}}]{Kena-Cohen2013}%
  \BibitemOpen
  \bibfield  {author} {\bibinfo {author} {\bibfnamefont {St\'{e}phane}\
  \bibnamefont {K\'{e}na-Cohen}}, \bibinfo {author} {\bibfnamefont {Stefan~A.}\
  \bibnamefont {Maier}}, \ and\ \bibinfo {author} {\bibfnamefont {Donal D.~C.}\
  \bibnamefont {Bradley}},\ }\bibfield  {title} {\enquote {\bibinfo {title}
  {{Ultrastrongly Coupled Exciton-Polaritons in Metal-Clad Organic
  Semiconductor Microcavities}},}\ }\href {\doibase 10.1002/adom.201300256}
  {\bibfield  {journal} {\bibinfo  {journal} {Adv. Opt. Mater.}\ }\textbf
  {\bibinfo {volume} {1}},\ \bibinfo {pages} {827--833} (\bibinfo {year}
  {2013})}\BibitemShut {NoStop}%
\bibitem [{\citenamefont {T\"{o}rm\"{a}}\ and\ \citenamefont
  {Barnes}(2015)}]{Torma2015}%
  \BibitemOpen
  \bibfield  {author} {\bibinfo {author} {\bibfnamefont {P.}~\bibnamefont
  {T\"{o}rm\"{a}}}\ and\ \bibinfo {author} {\bibfnamefont {W.~L.}\ \bibnamefont
  {Barnes}},\ }\bibfield  {title} {\enquote {\bibinfo {title} {{Strong coupling
  between surface plasmon polaritons and emitters: a review}},}\ }\href
  {\doibase 10.1088/0034-4885/78/1/013901} {\bibfield  {journal} {\bibinfo
  {journal} {Rep. Prog. Phys.}\ }\textbf {\bibinfo {volume} {78}},\ \bibinfo
  {pages} {013901} (\bibinfo {year} {2015})}\BibitemShut {NoStop}%
\bibitem [{\citenamefont {K\'{e}na-Cohen}\ \emph {et~al.}(2008)\citenamefont
  {K\'{e}na-Cohen}, \citenamefont {Davan\c{c}o},\ and\ \citenamefont
  {Forrest}}]{Kena-Cohen2008}%
  \BibitemOpen
  \bibfield  {author} {\bibinfo {author} {\bibfnamefont {S.}~\bibnamefont
  {K\'{e}na-Cohen}}, \bibinfo {author} {\bibfnamefont {M.}~\bibnamefont
  {Davan\c{c}o}}, \ and\ \bibinfo {author} {\bibfnamefont {S.~R.}\ \bibnamefont
  {Forrest}},\ }\bibfield  {title} {\enquote {\bibinfo {title} {{Strong
  Exciton-Photon Coupling in an Organic Single Crystal Microcavity}},}\ }\href
  {\doibase 10.1103/PhysRevLett.101.116401} {\bibfield  {journal} {\bibinfo
  {journal} {Phys. Rev. Lett.}\ }\textbf {\bibinfo {volume} {101}},\ \bibinfo
  {pages} {116401} (\bibinfo {year} {2008})}\BibitemShut {NoStop}%
\bibitem [{\citenamefont {Bellessa}\ \emph {et~al.}(2004)\citenamefont
  {Bellessa}, \citenamefont {Bonnand}, \citenamefont {Plenet},\ and\
  \citenamefont {Mugnier}}]{Bellessa2004}%
  \BibitemOpen
  \bibfield  {author} {\bibinfo {author} {\bibfnamefont {J.}~\bibnamefont
  {Bellessa}}, \bibinfo {author} {\bibfnamefont {C.}~\bibnamefont {Bonnand}},
  \bibinfo {author} {\bibfnamefont {J.~C.}\ \bibnamefont {Plenet}}, \ and\
  \bibinfo {author} {\bibfnamefont {J.}~\bibnamefont {Mugnier}},\ }\bibfield
  {title} {\enquote {\bibinfo {title} {{Strong Coupling between Surface
  Plasmons and Excitons in an Organic Semiconductor}},}\ }\href {\doibase
  10.1103/PhysRevLett.93.036404} {\bibfield  {journal} {\bibinfo  {journal}
  {Phys. Rev. Lett.}\ }\textbf {\bibinfo {volume} {93}},\ \bibinfo {pages}
  {036404} (\bibinfo {year} {2004})}\BibitemShut {NoStop}%
\bibitem [{\citenamefont {Dintinger}\ \emph {et~al.}(2005)\citenamefont
  {Dintinger}, \citenamefont {Klein}, \citenamefont {Bustos}, \citenamefont
  {Barnes},\ and\ \citenamefont {Ebbesen}}]{Dintinger2005}%
  \BibitemOpen
  \bibfield  {author} {\bibinfo {author} {\bibfnamefont {J.}~\bibnamefont
  {Dintinger}}, \bibinfo {author} {\bibfnamefont {S.}~\bibnamefont {Klein}},
  \bibinfo {author} {\bibfnamefont {F.}~\bibnamefont {Bustos}}, \bibinfo
  {author} {\bibfnamefont {W.~L.}\ \bibnamefont {Barnes}}, \ and\ \bibinfo
  {author} {\bibfnamefont {T.~W.}\ \bibnamefont {Ebbesen}},\ }\bibfield
  {title} {\enquote {\bibinfo {title} {{Strong coupling between surface
  plasmon-polaritons and organic molecules in subwavelength hole arrays}},}\
  }\href {\doibase 10.1103/PhysRevB.71.035424} {\bibfield  {journal} {\bibinfo
  {journal} {Phys. Rev. B}\ }\textbf {\bibinfo {volume} {71}},\ \bibinfo
  {pages} {035424} (\bibinfo {year} {2005})}\BibitemShut {NoStop}%
\bibitem [{\citenamefont {Hakala}\ \emph {et~al.}(2009)\citenamefont {Hakala},
  \citenamefont {Toppari}, \citenamefont {Kuzyk}, \citenamefont {Pettersson},
  \citenamefont {Tikkanen}, \citenamefont {Kunttu},\ and\ \citenamefont
  {T\"{o}rm\"{a}}}]{Hakala2009}%
  \BibitemOpen
  \bibfield  {author} {\bibinfo {author} {\bibfnamefont {T.~K.}\ \bibnamefont
  {Hakala}}, \bibinfo {author} {\bibfnamefont {J.~J.}\ \bibnamefont {Toppari}},
  \bibinfo {author} {\bibfnamefont {A.}~\bibnamefont {Kuzyk}}, \bibinfo
  {author} {\bibfnamefont {M.}~\bibnamefont {Pettersson}}, \bibinfo {author}
  {\bibfnamefont {H.}~\bibnamefont {Tikkanen}}, \bibinfo {author}
  {\bibfnamefont {H.}~\bibnamefont {Kunttu}}, \ and\ \bibinfo {author}
  {\bibfnamefont {P.}~\bibnamefont {T\"{o}rm\"{a}}},\ }\bibfield  {title}
  {\enquote {\bibinfo {title} {{Vacuum Rabi Splitting and Strong-Coupling
  Dynamics for Surface-Plasmon Polaritons and Rhodamine 6G Molecules}},}\
  }\href {\doibase 10.1103/PhysRevLett.103.053602} {\bibfield  {journal}
  {\bibinfo  {journal} {Phys. Rev. Lett.}\ }\textbf {\bibinfo {volume} {103}},\
  \bibinfo {pages} {053602} (\bibinfo {year} {2009})}\BibitemShut {NoStop}%
\bibitem [{\citenamefont {Vasa}\ \emph {et~al.}(2010)\citenamefont {Vasa},
  \citenamefont {Pomraenke}, \citenamefont {Cirmi}, \citenamefont {{De Re}},
  \citenamefont {Wang}, \citenamefont {Schwieger}, \citenamefont {Leipold},
  \citenamefont {Runge}, \citenamefont {Cerullo},\ and\ \citenamefont
  {Lienau}}]{Vasa2010}%
  \BibitemOpen
  \bibfield  {author} {\bibinfo {author} {\bibfnamefont {P.}~\bibnamefont
  {Vasa}}, \bibinfo {author} {\bibfnamefont {R.}~\bibnamefont {Pomraenke}},
  \bibinfo {author} {\bibfnamefont {G.}~\bibnamefont {Cirmi}}, \bibinfo
  {author} {\bibfnamefont {E.}~\bibnamefont {{De Re}}}, \bibinfo {author}
  {\bibfnamefont {W.}~\bibnamefont {Wang}}, \bibinfo {author} {\bibfnamefont
  {S.}~\bibnamefont {Schwieger}}, \bibinfo {author} {\bibfnamefont
  {D.}~\bibnamefont {Leipold}}, \bibinfo {author} {\bibfnamefont
  {E.}~\bibnamefont {Runge}}, \bibinfo {author} {\bibfnamefont
  {G.}~\bibnamefont {Cerullo}}, \ and\ \bibinfo {author} {\bibfnamefont
  {C.}~\bibnamefont {Lienau}},\ }\bibfield  {title} {\enquote {\bibinfo {title}
  {{Ultrafast manipulation of strong coupling in metal-molecular aggregate
  hybrid nanostructures}},}\ }\href {\doibase 10.1021/nn101973p} {\bibfield
  {journal} {\bibinfo  {journal} {ACS Nano}\ }\textbf {\bibinfo {volume} {4}},\
  \bibinfo {pages} {7559--65} (\bibinfo {year} {2010})}\BibitemShut {NoStop}%
\bibitem [{\citenamefont {Rodriguez}\ \emph {et~al.}(2013)\citenamefont
  {Rodriguez}, \citenamefont {Feist}, \citenamefont {Verschuuren},
  \citenamefont {{Garc\'{\i}a Vidal}},\ and\ \citenamefont {{G\'{o}mez
  Rivas}}}]{Rodriguez2013}%
  \BibitemOpen
  \bibfield  {author} {\bibinfo {author} {\bibfnamefont {S.~R.~K.}\
  \bibnamefont {Rodriguez}}, \bibinfo {author} {\bibfnamefont {J.}~\bibnamefont
  {Feist}}, \bibinfo {author} {\bibfnamefont {M.~A.}\ \bibnamefont
  {Verschuuren}}, \bibinfo {author} {\bibfnamefont {F.~J.}\ \bibnamefont
  {{Garc\'{\i}a Vidal}}}, \ and\ \bibinfo {author} {\bibfnamefont
  {J.}~\bibnamefont {{G\'{o}mez Rivas}}},\ }\bibfield  {title} {\enquote
  {\bibinfo {title} {{Thermalization and Cooling of Plasmon-Exciton Polaritons:
  Towards Quantum Condensation}},}\ }\href {\doibase
  10.1103/PhysRevLett.111.166802} {\bibfield  {journal} {\bibinfo  {journal}
  {Phys. Rev. Lett.}\ }\textbf {\bibinfo {volume} {111}},\ \bibinfo {pages}
  {166802} (\bibinfo {year} {2013})}\BibitemShut {NoStop}%
\bibitem [{\citenamefont {V\"{a}kev\"{a}inen}\ \emph
  {et~al.}(2014)\citenamefont {V\"{a}kev\"{a}inen}, \citenamefont {Moerland},
  \citenamefont {Rekola}, \citenamefont {Eskelinen}, \citenamefont
  {Martikainen}, \citenamefont {Kim},\ and\ \citenamefont
  {T\"{o}rm\"{a}}}]{Vakevainen2014}%
  \BibitemOpen
  \bibfield  {author} {\bibinfo {author} {\bibfnamefont {A.~I.}\ \bibnamefont
  {V\"{a}kev\"{a}inen}}, \bibinfo {author} {\bibfnamefont {R.~J.}\ \bibnamefont
  {Moerland}}, \bibinfo {author} {\bibfnamefont {H.~T.}\ \bibnamefont
  {Rekola}}, \bibinfo {author} {\bibfnamefont {A.-P.}\ \bibnamefont
  {Eskelinen}}, \bibinfo {author} {\bibfnamefont {J.-P.}\ \bibnamefont
  {Martikainen}}, \bibinfo {author} {\bibfnamefont {D.-H.}\ \bibnamefont
  {Kim}}, \ and\ \bibinfo {author} {\bibfnamefont {P.}~\bibnamefont
  {T\"{o}rm\"{a}}},\ }\bibfield  {title} {\enquote {\bibinfo {title}
  {{Plasmonic Surface Lattice Resonances at the Strong Coupling Regime}},}\
  }\href {\doibase 10.1021/nl4035219} {\bibfield  {journal} {\bibinfo
  {journal} {Nano Lett.}\ }\textbf {\bibinfo {volume} {14}},\ \bibinfo {pages}
  {1721--7} (\bibinfo {year} {2014})}\BibitemShut {NoStop}%
\bibitem [{\citenamefont {Baudrion}\ \emph {et~al.}(2013)\citenamefont
  {Baudrion}, \citenamefont {Perron}, \citenamefont {Veltri}, \citenamefont
  {Bouhelier}, \citenamefont {Adam},\ and\ \citenamefont
  {Bachelot}}]{Baudrion2013}%
  \BibitemOpen
  \bibfield  {author} {\bibinfo {author} {\bibfnamefont {Anne~Laure}\
  \bibnamefont {Baudrion}}, \bibinfo {author} {\bibfnamefont {Antoine}\
  \bibnamefont {Perron}}, \bibinfo {author} {\bibfnamefont {Alessandro}\
  \bibnamefont {Veltri}}, \bibinfo {author} {\bibfnamefont {Alexandre}\
  \bibnamefont {Bouhelier}}, \bibinfo {author} {\bibfnamefont {Pierre~Michel}\
  \bibnamefont {Adam}}, \ and\ \bibinfo {author} {\bibfnamefont {Renaud}\
  \bibnamefont {Bachelot}},\ }\bibfield  {title} {\enquote {\bibinfo {title}
  {{Reversible strong coupling in silver nanoparticle arrays using photochromic
  molecules}},}\ }\href {\doibase 10.1021/nl3040948} {\bibfield  {journal}
  {\bibinfo  {journal} {Nano Lett.}\ }\textbf {\bibinfo {volume} {13}},\
  \bibinfo {pages} {282--286} (\bibinfo {year} {2013})}\BibitemShut {NoStop}%
\bibitem [{\citenamefont {Zengin}\ \emph {et~al.}(2015)\citenamefont {Zengin},
  \citenamefont {Wers\"{a}ll}, \citenamefont {Nilsson}, \citenamefont
  {Antosiewicz}, \citenamefont {K\"{a}ll},\ and\ \citenamefont
  {Shegai}}]{Zengin2015}%
  \BibitemOpen
  \bibfield  {author} {\bibinfo {author} {\bibfnamefont {G\"{u}lis}\
  \bibnamefont {Zengin}}, \bibinfo {author} {\bibfnamefont {Martin}\
  \bibnamefont {Wers\"{a}ll}}, \bibinfo {author} {\bibfnamefont {Sara}\
  \bibnamefont {Nilsson}}, \bibinfo {author} {\bibfnamefont {Tomasz~J.}\
  \bibnamefont {Antosiewicz}}, \bibinfo {author} {\bibfnamefont {Mikael}\
  \bibnamefont {K\"{a}ll}}, \ and\ \bibinfo {author} {\bibfnamefont {Timur}\
  \bibnamefont {Shegai}},\ }\bibfield  {title} {\enquote {\bibinfo {title}
  {{Realizing Strong Light-Matter Interactions between Single-Nanoparticle
  Plasmons and Molecular Excitons at Ambient Conditions}},}\ }\href {\doibase
  10.1103/PhysRevLett.114.157401} {\bibfield  {journal} {\bibinfo  {journal}
  {Phys. Rev. Lett.}\ }\textbf {\bibinfo {volume} {114}},\ \bibinfo {pages}
  {157401} (\bibinfo {year} {2015})}\BibitemShut {NoStop}%
\bibitem [{\citenamefont {Hutchison}\ \emph {et~al.}(2012)\citenamefont
  {Hutchison}, \citenamefont {Schwartz}, \citenamefont {Genet}, \citenamefont
  {Devaux},\ and\ \citenamefont {Ebbesen}}]{Hutchison2012}%
  \BibitemOpen
  \bibfield  {author} {\bibinfo {author} {\bibfnamefont {James~A.}\
  \bibnamefont {Hutchison}}, \bibinfo {author} {\bibfnamefont {Tal}\
  \bibnamefont {Schwartz}}, \bibinfo {author} {\bibfnamefont {Cyriaque}\
  \bibnamefont {Genet}}, \bibinfo {author} {\bibfnamefont {Elo\"{\i}se}\
  \bibnamefont {Devaux}}, \ and\ \bibinfo {author} {\bibfnamefont {Thomas~W.}\
  \bibnamefont {Ebbesen}},\ }\bibfield  {title} {\enquote {\bibinfo {title}
  {{Modifying Chemical Landscapes by Coupling to Vacuum Fields}},}\ }\href
  {\doibase 10.1002/ange.201107033} {\bibfield  {journal} {\bibinfo  {journal}
  {Angew. Chemie}\ }\textbf {\bibinfo {volume} {124}},\ \bibinfo {pages}
  {1624--1628} (\bibinfo {year} {2012})}\BibitemShut {NoStop}%
\bibitem [{\citenamefont {Hutchison}\ \emph {et~al.}(2013)\citenamefont
  {Hutchison}, \citenamefont {Liscio}, \citenamefont {Schwartz}, \citenamefont
  {Canaguier-Durand}, \citenamefont {Genet}, \citenamefont {Palermo},
  \citenamefont {Samor\`{\i}},\ and\ \citenamefont {Ebbesen}}]{Hutchison2013}%
  \BibitemOpen
  \bibfield  {author} {\bibinfo {author} {\bibfnamefont {James~A.}\
  \bibnamefont {Hutchison}}, \bibinfo {author} {\bibfnamefont {Andrea}\
  \bibnamefont {Liscio}}, \bibinfo {author} {\bibfnamefont {Tal}\ \bibnamefont
  {Schwartz}}, \bibinfo {author} {\bibfnamefont {Antoine}\ \bibnamefont
  {Canaguier-Durand}}, \bibinfo {author} {\bibfnamefont {Cyriaque}\
  \bibnamefont {Genet}}, \bibinfo {author} {\bibfnamefont {Vincenzo}\
  \bibnamefont {Palermo}}, \bibinfo {author} {\bibfnamefont {Paolo}\
  \bibnamefont {Samor\`{\i}}}, \ and\ \bibinfo {author} {\bibfnamefont
  {Thomas~W.}\ \bibnamefont {Ebbesen}},\ }\bibfield  {title} {\enquote
  {\bibinfo {title} {{Tuning the work-function via strong coupling}},}\ }\href
  {\doibase 10.1002/adma.201203682} {\bibfield  {journal} {\bibinfo  {journal}
  {Adv. Mater.}\ }\textbf {\bibinfo {volume} {25}},\ \bibinfo {pages} {2481--5}
  (\bibinfo {year} {2013})}\BibitemShut {NoStop}%
\bibitem [{\citenamefont {Wang}\ \emph {et~al.}(2014)\citenamefont {Wang},
  \citenamefont {Mika}, \citenamefont {Hutchison}, \citenamefont {Genet},
  \citenamefont {Jouaiti}, \citenamefont {Hosseini},\ and\ \citenamefont
  {Ebbesen}}]{Wang2014b}%
  \BibitemOpen
  \bibfield  {author} {\bibinfo {author} {\bibfnamefont {Shaojun}\ \bibnamefont
  {Wang}}, \bibinfo {author} {\bibfnamefont {Arkadiusz}\ \bibnamefont {Mika}},
  \bibinfo {author} {\bibfnamefont {James~A.}\ \bibnamefont {Hutchison}},
  \bibinfo {author} {\bibfnamefont {Cyriaque}\ \bibnamefont {Genet}}, \bibinfo
  {author} {\bibfnamefont {Abdelaziz}\ \bibnamefont {Jouaiti}}, \bibinfo
  {author} {\bibfnamefont {Mir~Wais}\ \bibnamefont {Hosseini}}, \ and\ \bibinfo
  {author} {\bibfnamefont {Thomas~W.}\ \bibnamefont {Ebbesen}},\ }\bibfield
  {title} {\enquote {\bibinfo {title} {{Phase transition of a perovskite
  strongly coupled to the vacuum field}},}\ }\href {\doibase
  10.1039/c4nr01971g} {\bibfield  {journal} {\bibinfo  {journal} {Nanoscale}\
  }\textbf {\bibinfo {volume} {6}},\ \bibinfo {pages} {7243--8} (\bibinfo
  {year} {2014})}\BibitemShut {NoStop}%
\bibitem [{\citenamefont {Carusotto}\ and\ \citenamefont
  {Ciuti}(2013)}]{Carusotto2013}%
  \BibitemOpen
  \bibfield  {author} {\bibinfo {author} {\bibfnamefont {Iacopo}\ \bibnamefont
  {Carusotto}}\ and\ \bibinfo {author} {\bibfnamefont {Cristiano}\ \bibnamefont
  {Ciuti}},\ }\bibfield  {title} {\enquote {\bibinfo {title} {{Quantum fluids
  of light}},}\ }\href {\doibase 10.1103/RevModPhys.85.299} {\bibfield
  {journal} {\bibinfo  {journal} {Rev. Mod. Phys.}\ }\textbf {\bibinfo {volume}
  {85}},\ \bibinfo {pages} {299--366} (\bibinfo {year} {2013})}\BibitemShut
  {NoStop}%
\bibitem [{\citenamefont {Michetti}\ \emph {et~al.}(2015)\citenamefont
  {Michetti}, \citenamefont {Mazza},\ and\ \citenamefont {{La
  Rocca}}}]{Michetti2015}%
  \BibitemOpen
  \bibfield  {author} {\bibinfo {author} {\bibfnamefont {Paolo}\ \bibnamefont
  {Michetti}}, \bibinfo {author} {\bibfnamefont {Leonardo}\ \bibnamefont
  {Mazza}}, \ and\ \bibinfo {author} {\bibfnamefont {Giuseppe~C.}\ \bibnamefont
  {{La Rocca}}},\ }\bibfield  {title} {\enquote {\bibinfo {title} {{Strongly
  Coupled Organic Microcavities}},}\ }in\ \href {\doibase
  10.1007/978-3-662-45082-6\_2} {\emph {\bibinfo {booktitle} {Organic
  Nanophotonics}}},\ \bibinfo {series and number} {Nano-Optics and
  Nanophotonics},\ \bibinfo {editor} {edited by\ \bibinfo {editor}
  {\bibfnamefont {Yong~Sheng}\ \bibnamefont {Zhao}}}\ (\bibinfo  {publisher}
  {Springer},\ \bibinfo {address} {Berlin, Heidelberg},\ \bibinfo {year}
  {2015})\ pp.\ \bibinfo {pages} {39--68}\BibitemShut {NoStop}%
\bibitem [{\citenamefont {Gonz\'{a}lez-Tudela}\ \emph
  {et~al.}(2013)\citenamefont {Gonz\'{a}lez-Tudela}, \citenamefont {Huidobro},
  \citenamefont {Mart\'{\i}n-Moreno}, \citenamefont {Tejedor},\ and\
  \citenamefont {Garc\'{\i}a-Vidal}}]{Gonzalez-Tudela2013}%
  \BibitemOpen
  \bibfield  {author} {\bibinfo {author} {\bibfnamefont {A.}~\bibnamefont
  {Gonz\'{a}lez-Tudela}}, \bibinfo {author} {\bibfnamefont {P.~A.}\
  \bibnamefont {Huidobro}}, \bibinfo {author} {\bibfnamefont {L.}~\bibnamefont
  {Mart\'{\i}n-Moreno}}, \bibinfo {author} {\bibfnamefont {C.}~\bibnamefont
  {Tejedor}}, \ and\ \bibinfo {author} {\bibfnamefont {F.~J.}\ \bibnamefont
  {Garc\'{\i}a-Vidal}},\ }\bibfield  {title} {\enquote {\bibinfo {title}
  {{Theory of Strong Coupling between Quantum Emitters and Propagating Surface
  Plasmons}},}\ }\href {\doibase 10.1103/PhysRevLett.110.126801} {\bibfield
  {journal} {\bibinfo  {journal} {Phys. Rev. Lett.}\ }\textbf {\bibinfo
  {volume} {110}},\ \bibinfo {pages} {126801} (\bibinfo {year}
  {2013})}\BibitemShut {NoStop}%
\bibitem [{\citenamefont {Mazza}\ \emph {et~al.}(2013)\citenamefont {Mazza},
  \citenamefont {K\'{e}na-Cohen}, \citenamefont {Michetti},\ and\ \citenamefont
  {{La Rocca}}}]{Mazza2013}%
  \BibitemOpen
  \bibfield  {author} {\bibinfo {author} {\bibfnamefont {L.}~\bibnamefont
  {Mazza}}, \bibinfo {author} {\bibfnamefont {S.}~\bibnamefont
  {K\'{e}na-Cohen}}, \bibinfo {author} {\bibfnamefont {P.}~\bibnamefont
  {Michetti}}, \ and\ \bibinfo {author} {\bibfnamefont {G.~C.}\ \bibnamefont
  {{La Rocca}}},\ }\bibfield  {title} {\enquote {\bibinfo {title} {{Microscopic
  theory of polariton lasing via vibronically assisted scattering}},}\ }\href
  {\doibase 10.1103/PhysRevB.88.075321} {\bibfield  {journal} {\bibinfo
  {journal} {Phys. Rev. B}\ }\textbf {\bibinfo {volume} {88}},\ \bibinfo
  {pages} {075321} (\bibinfo {year} {2013})}\BibitemShut {NoStop}%
\bibitem [{\citenamefont {\'{C}wik}\ \emph {et~al.}(2014)\citenamefont
  {\'{C}wik}, \citenamefont {Reja}, \citenamefont {Littlewood},\ and\
  \citenamefont {Keeling}}]{Cwik2014}%
  \BibitemOpen
  \bibfield  {author} {\bibinfo {author} {\bibfnamefont {Justyna~A.}\
  \bibnamefont {\'{C}wik}}, \bibinfo {author} {\bibfnamefont {Sahinur}\
  \bibnamefont {Reja}}, \bibinfo {author} {\bibfnamefont {Peter~B.}\
  \bibnamefont {Littlewood}}, \ and\ \bibinfo {author} {\bibfnamefont
  {Jonathan}\ \bibnamefont {Keeling}},\ }\bibfield  {title} {\enquote {\bibinfo
  {title} {{Polariton condensation with saturable molecules dressed by
  vibrational modes}},}\ }\href {\doibase 10.1209/0295-5075/105/47009}
  {\bibfield  {journal} {\bibinfo  {journal} {EPL}\ }\textbf {\bibinfo {volume}
  {105}},\ \bibinfo {pages} {47009} (\bibinfo {year} {2014})}\BibitemShut
  {NoStop}%
\bibitem [{\citenamefont {Born}\ and\ \citenamefont
  {Oppenheimer}(1927)}]{Born1927}%
  \BibitemOpen
  \bibfield  {author} {\bibinfo {author} {\bibfnamefont {M.}~\bibnamefont
  {Born}}\ and\ \bibinfo {author} {\bibfnamefont {R.}~\bibnamefont
  {Oppenheimer}},\ }\bibfield  {title} {\enquote {\bibinfo {title} {{Zur
  Quantentheorie der Molekeln}},}\ }\href {\doibase 10.1002/andp.19273892002}
  {\bibfield  {journal} {\bibinfo  {journal} {Ann. Phys.}\ }\textbf {\bibinfo
  {volume} {20}},\ \bibinfo {pages} {457--484} (\bibinfo {year}
  {1927})}\BibitemShut {NoStop}%
\bibitem [{\citenamefont {Tully}(2000)}]{Tully2000}%
  \BibitemOpen
  \bibfield  {author} {\bibinfo {author} {\bibfnamefont {John~C.}\ \bibnamefont
  {Tully}},\ }\bibfield  {title} {\enquote {\bibinfo {title} {{Perspective on
  "Zur Quantentheorie der Molekeln"}},}\ }\href {\doibase
  10.1007/s002149900049} {\bibfield  {journal} {\bibinfo  {journal} {Theor.
  Chem. Acc.}\ }\textbf {\bibinfo {volume} {103}},\ \bibinfo {pages} {173--176}
  (\bibinfo {year} {2000})}\BibitemShut {NoStop}%
\bibitem [{\citenamefont {Houdr\'{e}}\ \emph {et~al.}(1996)\citenamefont
  {Houdr\'{e}}, \citenamefont {Stanley},\ and\ \citenamefont
  {Ilegems}}]{Houdre1996}%
  \BibitemOpen
  \bibfield  {author} {\bibinfo {author} {\bibfnamefont {R.}~\bibnamefont
  {Houdr\'{e}}}, \bibinfo {author} {\bibfnamefont {R.~P.}\ \bibnamefont
  {Stanley}}, \ and\ \bibinfo {author} {\bibfnamefont {M.}~\bibnamefont
  {Ilegems}},\ }\bibfield  {title} {\enquote {\bibinfo {title} {{Vacuum-field
  Rabi splitting in the presence of inhomogeneous broadening: Resolution of a
  homogeneous linewidth in an inhomogeneously broadened system}},}\ }\href
  {\doibase 10.1103/PhysRevA.53.2711} {\bibfield  {journal} {\bibinfo
  {journal} {Phys. Rev. A}\ }\textbf {\bibinfo {volume} {53}},\ \bibinfo
  {pages} {2711--2715} (\bibinfo {year} {1996})}\BibitemShut {NoStop}%
\bibitem [{\citenamefont {Agranovich}\ \emph {et~al.}(2011)\citenamefont
  {Agranovich}, \citenamefont {Gartstein},\ and\ \citenamefont
  {Litinskaya}}]{Agranovich2011}%
  \BibitemOpen
  \bibfield  {author} {\bibinfo {author} {\bibfnamefont {V.~M.}\ \bibnamefont
  {Agranovich}}, \bibinfo {author} {\bibfnamefont {Yu.~N.}\ \bibnamefont
  {Gartstein}}, \ and\ \bibinfo {author} {\bibfnamefont {M.}~\bibnamefont
  {Litinskaya}},\ }\bibfield  {title} {\enquote {\bibinfo {title} {{Hybrid
  resonant organic-inorganic nanostructures for optoelectronic
  applications}},}\ }\href {\doibase 10.1021/cr100156x} {\bibfield  {journal}
  {\bibinfo  {journal} {Chem. Rev.}\ }\textbf {\bibinfo {volume} {111}},\
  \bibinfo {pages} {5179--5214} (\bibinfo {year} {2011})}\BibitemShut {NoStop}%
\bibitem [{\citenamefont {{De Liberato}}\ \emph {et~al.}(2007)\citenamefont
  {{De Liberato}}, \citenamefont {Ciuti},\ and\ \citenamefont
  {Carusotto}}]{DeLiberato2007}%
  \BibitemOpen
  \bibfield  {author} {\bibinfo {author} {\bibfnamefont {Simone}\ \bibnamefont
  {{De Liberato}}}, \bibinfo {author} {\bibfnamefont {Cristiano}\ \bibnamefont
  {Ciuti}}, \ and\ \bibinfo {author} {\bibfnamefont {Iacopo}\ \bibnamefont
  {Carusotto}},\ }\bibfield  {title} {\enquote {\bibinfo {title} {{Quantum
  Vacuum Radiation Spectra from a Semiconductor Microcavity with a
  Time-Modulated Vacuum Rabi Frequency}},}\ }\href {\doibase
  10.1103/PhysRevLett.98.103602} {\bibfield  {journal} {\bibinfo  {journal}
  {Phys. Rev. Lett.}\ }\textbf {\bibinfo {volume} {98}},\ \bibinfo {pages}
  {103602} (\bibinfo {year} {2007})}\BibitemShut {NoStop}%
\bibitem [{\citenamefont {George}\ \emph {et~al.}(2015)\citenamefont {George},
  \citenamefont {Wang}, \citenamefont {Chervy}, \citenamefont
  {Canaguier-Durand}, \citenamefont {Schaeffer}, \citenamefont {Lehn},
  \citenamefont {Hutchison}, \citenamefont {Genet},\ and\ \citenamefont
  {Ebbesen}}]{George2015}%
  \BibitemOpen
  \bibfield  {author} {\bibinfo {author} {\bibfnamefont {Jino}\ \bibnamefont
  {George}}, \bibinfo {author} {\bibfnamefont {Shaojun}\ \bibnamefont {Wang}},
  \bibinfo {author} {\bibfnamefont {Thibault}\ \bibnamefont {Chervy}}, \bibinfo
  {author} {\bibfnamefont {Antoine}\ \bibnamefont {Canaguier-Durand}}, \bibinfo
  {author} {\bibfnamefont {Gael}\ \bibnamefont {Schaeffer}}, \bibinfo {author}
  {\bibfnamefont {Jean-Marie}\ \bibnamefont {Lehn}}, \bibinfo {author}
  {\bibfnamefont {James~A.}\ \bibnamefont {Hutchison}}, \bibinfo {author}
  {\bibfnamefont {Cyriaque}\ \bibnamefont {Genet}}, \ and\ \bibinfo {author}
  {\bibfnamefont {Thomas~W.}\ \bibnamefont {Ebbesen}},\ }\bibfield  {title}
  {\enquote {\bibinfo {title} {{Ultra-strong coupling of molecular materials:
  spectroscopy and dynamics}},}\ }\href {\doibase 10.1039/C4FD00197D}
  {\bibfield  {journal} {\bibinfo  {journal} {Faraday Discuss.}\ }\textbf
  {\bibinfo {volume} {178}},\ \bibinfo {pages} {281--294} (\bibinfo {year}
  {2015})}\BibitemShut {NoStop}%
\bibitem [{\citenamefont {May}\ and\ \citenamefont {K\"{u}hn}(2011)}]{May2011}%
  \BibitemOpen
  \bibfield  {author} {\bibinfo {author} {\bibfnamefont {Volkhard}\
  \bibnamefont {May}}\ and\ \bibinfo {author} {\bibfnamefont {Oliver}\
  \bibnamefont {K\"{u}hn}},\ }\href {\doibase 10.1002/9783527633791} {\emph
  {\bibinfo {title} {{Charge and Energy Transfer Dynamics in Molecular
  Systems}}}}\ (\bibinfo  {publisher} {Wiley-VCH Verlag GmbH \& Co. KGaA},\
  \bibinfo {address} {Weinheim, Germany},\ \bibinfo {year} {2011})\BibitemShut
  {NoStop}%
\bibitem [{\citenamefont {Kim}\ \emph {et~al.}(2015)\citenamefont {Kim},
  \citenamefont {Sim}, \citenamefont {Yoon}, \citenamefont {Gong},
  \citenamefont {Ahn}, \citenamefont {Cho},\ and\ \citenamefont
  {Lee}}]{Kim2015}%
  \BibitemOpen
  \bibfield  {author} {\bibinfo {author} {\bibfnamefont {Myung-Ki}\
  \bibnamefont {Kim}}, \bibinfo {author} {\bibfnamefont {Hongchul}\
  \bibnamefont {Sim}}, \bibinfo {author} {\bibfnamefont {Seung~Ju}\
  \bibnamefont {Yoon}}, \bibinfo {author} {\bibfnamefont {Su-Hyun}\
  \bibnamefont {Gong}}, \bibinfo {author} {\bibfnamefont {Chi~Won}\
  \bibnamefont {Ahn}}, \bibinfo {author} {\bibfnamefont {Yong-Hoon}\
  \bibnamefont {Cho}}, \ and\ \bibinfo {author} {\bibfnamefont {Yong-Hee}\
  \bibnamefont {Lee}},\ }\bibfield  {title} {\enquote {\bibinfo {title}
  {{Squeezing Photons into a Point-Like Space}},}\ }\href {\doibase
  10.1021/acs.nanolett.5b01204} {\bibfield  {journal} {\bibinfo  {journal}
  {Nano Lett.}\ } (\bibinfo {year} {2015}),\
  10.1021/acs.nanolett.5b01204}\BibitemShut {NoStop}%
\bibitem [{\citenamefont {Moll}\ \emph {et~al.}(1995)\citenamefont {Moll},
  \citenamefont {Daehne}, \citenamefont {Durrant},\ and\ \citenamefont
  {Wiersma}}]{Moll1995}%
  \BibitemOpen
  \bibfield  {author} {\bibinfo {author} {\bibfnamefont {Johannes}\
  \bibnamefont {Moll}}, \bibinfo {author} {\bibfnamefont {Siegfried}\
  \bibnamefont {Daehne}}, \bibinfo {author} {\bibfnamefont {James~R.}\
  \bibnamefont {Durrant}}, \ and\ \bibinfo {author} {\bibfnamefont {Douwe~A.}\
  \bibnamefont {Wiersma}},\ }\bibfield  {title} {\enquote {\bibinfo {title}
  {{Optical dynamics of excitons in J aggregates of a carbocyanine dye}},}\
  }\href {\doibase 10.1063/1.1703017} {\bibfield  {journal} {\bibinfo
  {journal} {J. Chem. Phys.}\ }\textbf {\bibinfo {volume} {102}},\ \bibinfo
  {pages} {6362} (\bibinfo {year} {1995})}\BibitemShut {NoStop}%
\bibitem [{\citenamefont {Rescigno}\ and\ \citenamefont
  {McKoy}(1975)}]{Rescigno1975}%
  \BibitemOpen
  \bibfield  {author} {\bibinfo {author} {\bibfnamefont {Thomas~N.}\
  \bibnamefont {Rescigno}}\ and\ \bibinfo {author} {\bibfnamefont {Vincent}\
  \bibnamefont {McKoy}},\ }\bibfield  {title} {\enquote {\bibinfo {title}
  {{Rigorous method for computing photoabsorption cross sections from a
  basis-set expansion}},}\ }\href {\doibase 10.1103/PhysRevA.12.522} {\bibfield
   {journal} {\bibinfo  {journal} {Phys. Rev. A}\ }\textbf {\bibinfo {volume}
  {12}},\ \bibinfo {pages} {522--525} (\bibinfo {year} {1975})}\BibitemShut
  {NoStop}%
\bibitem [{\citenamefont {Bonin}\ and\ \citenamefont
  {Kresin}(1997)}]{BonKre1997}%
  \BibitemOpen
  \bibfield  {author} {\bibinfo {author} {\bibfnamefont {Keith~D.}\
  \bibnamefont {Bonin}}\ and\ \bibinfo {author} {\bibfnamefont {Vitaly~V.}\
  \bibnamefont {Kresin}},\ }\href {\doibase 10.1142/9789814261272} {\emph
  {\bibinfo {title} {{Electric-Dipole Polarizabilities Of Atoms, Molecules, And
  Clusters}}}}\ (\bibinfo  {publisher} {World Scientific Publishing Co. Pte.
  Ltd.},\ \bibinfo {year} {1997})\BibitemShut {NoStop}%
\bibitem [{\citenamefont {Shimanouchi}(1972)}]{Shimanouchi1972}%
  \BibitemOpen
  \bibfield  {author} {\bibinfo {author} {\bibfnamefont {Takehiko}\
  \bibnamefont {Shimanouchi}},\ }\bibfield  {title} {\enquote {\bibinfo {title}
  {{Tables of Molecular Vibrational Frequencies. Consolidated Volume I}},}\
  }\href {http://www.nist.gov/data/nsrds/NSRDS-NBS-39.pdf} {\bibfield
  {journal} {\bibinfo  {journal} {NSRDS-NBS}\ }\textbf {\bibinfo {volume}
  {39}},\ \bibinfo {pages} {161p.} (\bibinfo {year} {1972})}\BibitemShut
  {NoStop}%
\bibitem [{\citenamefont {Dicke}(1954)}]{Dicke1954}%
  \BibitemOpen
  \bibfield  {author} {\bibinfo {author} {\bibfnamefont {R.}~\bibnamefont
  {Dicke}},\ }\bibfield  {title} {\enquote {\bibinfo {title} {{Coherence in
  Spontaneous Radiation Processes}},}\ }\href {\doibase 10.1103/PhysRev.93.99}
  {\bibfield  {journal} {\bibinfo  {journal} {Phys. Rev.}\ }\textbf {\bibinfo
  {volume} {93}},\ \bibinfo {pages} {99--110} (\bibinfo {year}
  {1954})}\BibitemShut {NoStop}%
\bibitem [{\citenamefont {Jaynes}\ and\ \citenamefont
  {Cummings}(1963)}]{Jaynes1963}%
  \BibitemOpen
  \bibfield  {author} {\bibinfo {author} {\bibfnamefont {E.~T.}\ \bibnamefont
  {Jaynes}}\ and\ \bibinfo {author} {\bibfnamefont {F.~W.}\ \bibnamefont
  {Cummings}},\ }\bibfield  {title} {\enquote {\bibinfo {title} {{Comparison of
  quantum and semiclassical radiation theories with application to the beam
  maser}},}\ }\href {\doibase 10.1109/PROC.1963.1664} {\bibfield  {journal}
  {\bibinfo  {journal} {Proc. IEEE}\ }\textbf {\bibinfo {volume} {51}},\
  \bibinfo {pages} {89--109} (\bibinfo {year} {1963})}\BibitemShut {NoStop}%
\bibitem [{\citenamefont {Tavis}\ and\ \citenamefont
  {Cummings}(1967)}]{Tavis1967}%
  \BibitemOpen
  \bibfield  {author} {\bibinfo {author} {\bibfnamefont {M.}~\bibnamefont
  {Tavis}}\ and\ \bibinfo {author} {\bibfnamefont {F.~W.}\ \bibnamefont
  {Cummings}},\ }\bibfield  {title} {\enquote {\bibinfo {title} {{The exact
  solution of N two level systems interacting with a single mode, quantized
  radiation field}},}\ }\href {\doibase 10.1016/0375-9601(67)90957-7}
  {\bibfield  {journal} {\bibinfo  {journal} {Phys. Lett. A}\ }\textbf
  {\bibinfo {volume} {25}},\ \bibinfo {pages} {714--715} (\bibinfo {year}
  {1967})}\BibitemShut {NoStop}%
\bibitem [{\citenamefont {K\"{o}nig}\ \emph {et~al.}(2012)\citenamefont
  {K\"{o}nig}, \citenamefont {van Bokhoven}, \citenamefont {Schildhauer},\ and\
  \citenamefont {Nachtegaal}}]{Konig2012}%
  \BibitemOpen
  \bibfield  {author} {\bibinfo {author} {\bibfnamefont {Christian F.~J.}\
  \bibnamefont {K\"{o}nig}}, \bibinfo {author} {\bibfnamefont {Jeroen~A.}\
  \bibnamefont {van Bokhoven}}, \bibinfo {author} {\bibfnamefont {Tilman~J.}\
  \bibnamefont {Schildhauer}}, \ and\ \bibinfo {author} {\bibfnamefont
  {Maarten}\ \bibnamefont {Nachtegaal}},\ }\bibfield  {title} {\enquote
  {\bibinfo {title} {{Quantitative analysis of modulated excitation X-ray
  absorption spectra: Enhanced precision of EXAFS fitting}},}\ }\href {\doibase
  10.1021/jp306022k} {\bibfield  {journal} {\bibinfo  {journal} {J. Phys. Chem.
  C}\ }\textbf {\bibinfo {volume} {116}},\ \bibinfo {pages} {19857--19866}
  (\bibinfo {year} {2012})}\BibitemShut {NoStop}%
\bibitem [{\citenamefont {Kozina}\ \emph {et~al.}(2014)\citenamefont {Kozina},
  \citenamefont {Hu}, \citenamefont {Wittenberg}, \citenamefont {Szilagyi},
  \citenamefont {Trigo}, \citenamefont {Miller}, \citenamefont {Uher},
  \citenamefont {Damodaran}, \citenamefont {Martin}, \citenamefont {Mehta},
  \citenamefont {Corbett}, \citenamefont {Safranek}, \citenamefont {Reis},\
  and\ \citenamefont {Lindenberg}}]{Kozina2014}%
  \BibitemOpen
  \bibfield  {author} {\bibinfo {author} {\bibfnamefont {M.}~\bibnamefont
  {Kozina}}, \bibinfo {author} {\bibfnamefont {T.}~\bibnamefont {Hu}}, \bibinfo
  {author} {\bibfnamefont {J.~S.}\ \bibnamefont {Wittenberg}}, \bibinfo
  {author} {\bibfnamefont {E.}~\bibnamefont {Szilagyi}}, \bibinfo {author}
  {\bibfnamefont {M.}~\bibnamefont {Trigo}}, \bibinfo {author} {\bibfnamefont
  {T.~A.}\ \bibnamefont {Miller}}, \bibinfo {author} {\bibfnamefont
  {C.}~\bibnamefont {Uher}}, \bibinfo {author} {\bibfnamefont {A.}~\bibnamefont
  {Damodaran}}, \bibinfo {author} {\bibfnamefont {L.}~\bibnamefont {Martin}},
  \bibinfo {author} {\bibfnamefont {A.}~\bibnamefont {Mehta}}, \bibinfo
  {author} {\bibfnamefont {J.}~\bibnamefont {Corbett}}, \bibinfo {author}
  {\bibfnamefont {J.}~\bibnamefont {Safranek}}, \bibinfo {author}
  {\bibfnamefont {D.~A.}\ \bibnamefont {Reis}}, \ and\ \bibinfo {author}
  {\bibfnamefont {A.~M.}\ \bibnamefont {Lindenberg}},\ }\bibfield  {title}
  {\enquote {\bibinfo {title} {{Measurement of transient atomic displacements
  in thin films with picosecond and femtometer resolution}},}\ }\href {\doibase
  10.1063/1.4875347} {\bibfield  {journal} {\bibinfo  {journal} {Struct. Dyn.}\
  }\textbf {\bibinfo {volume} {1}},\ \bibinfo {pages} {034301} (\bibinfo {year}
  {2014})}\BibitemShut {NoStop}%
\bibitem [{\citenamefont {Tokatly}(2013)}]{Tokatly2013}%
  \BibitemOpen
  \bibfield  {author} {\bibinfo {author} {\bibfnamefont {I.~V.}\ \bibnamefont
  {Tokatly}},\ }\bibfield  {title} {\enquote {\bibinfo {title} {{Time-dependent
  density functional theory for many-electron systems interacting with cavity
  photons}},}\ }\href {\doibase 10.1103/PhysRevLett.110.233001} {\bibfield
  {journal} {\bibinfo  {journal} {Phys. Rev. Lett.}\ }\textbf {\bibinfo
  {volume} {110}},\ \bibinfo {pages} {233001} (\bibinfo {year}
  {2013})}\BibitemShut {NoStop}%
\bibitem [{\citenamefont {Ruggenthaler}\ \emph {et~al.}(2014)\citenamefont
  {Ruggenthaler}, \citenamefont {Flick}, \citenamefont {Pellegrini},
  \citenamefont {Appel}, \citenamefont {Tokatly},\ and\ \citenamefont
  {Rubio}}]{Ruggenthaler2014}%
  \BibitemOpen
  \bibfield  {author} {\bibinfo {author} {\bibfnamefont {Michael}\ \bibnamefont
  {Ruggenthaler}}, \bibinfo {author} {\bibfnamefont {Johannes}\ \bibnamefont
  {Flick}}, \bibinfo {author} {\bibfnamefont {Camilla}\ \bibnamefont
  {Pellegrini}}, \bibinfo {author} {\bibfnamefont {Heiko}\ \bibnamefont
  {Appel}}, \bibinfo {author} {\bibfnamefont {Ilya~V.}\ \bibnamefont
  {Tokatly}}, \ and\ \bibinfo {author} {\bibfnamefont {Angel}\ \bibnamefont
  {Rubio}},\ }\bibfield  {title} {\enquote {\bibinfo {title}
  {{Quantum-electrodynamical density-functional theory: Bridging quantum optics
  and electronic-structure theory}},}\ }\href {\doibase
  10.1103/PhysRevA.90.012508} {\bibfield  {journal} {\bibinfo  {journal} {Phys.
  Rev. A}\ }\textbf {\bibinfo {volume} {90}},\ \bibinfo {pages} {012508}
  (\bibinfo {year} {2014})}\BibitemShut {NoStop}%
\bibitem [{\citenamefont {Leggett}\ \emph {et~al.}(1987)\citenamefont
  {Leggett}, \citenamefont {Chakravarty}, \citenamefont {Dorsey}, \citenamefont
  {Fisher}, \citenamefont {Garg},\ and\ \citenamefont {Zwerger}}]{Leggett1987}%
  \BibitemOpen
  \bibfield  {author} {\bibinfo {author} {\bibfnamefont {A.~J.}\ \bibnamefont
  {Leggett}}, \bibinfo {author} {\bibfnamefont {S.}~\bibnamefont
  {Chakravarty}}, \bibinfo {author} {\bibfnamefont {A.~T.}\ \bibnamefont
  {Dorsey}}, \bibinfo {author} {\bibfnamefont {Matthew P.~A.}\ \bibnamefont
  {Fisher}}, \bibinfo {author} {\bibfnamefont {Anupam}\ \bibnamefont {Garg}}, \
  and\ \bibinfo {author} {\bibfnamefont {W.}~\bibnamefont {Zwerger}},\
  }\bibfield  {title} {\enquote {\bibinfo {title} {{Dynamics of the dissipative
  two-state system}},}\ }\href {\doibase 10.1103/RevModPhys.59.1} {\bibfield
  {journal} {\bibinfo  {journal} {Rev. Mod. Phys.}\ }\textbf {\bibinfo {volume}
  {59}},\ \bibinfo {pages} {1--85} (\bibinfo {year} {1987})}\BibitemShut
  {NoStop}%
\bibitem [{\citenamefont {Coalson}\ \emph {et~al.}(1994)\citenamefont
  {Coalson}, \citenamefont {Evans},\ and\ \citenamefont
  {Nitzan}}]{Coalson1994}%
  \BibitemOpen
  \bibfield  {author} {\bibinfo {author} {\bibfnamefont {Rob~D.}\ \bibnamefont
  {Coalson}}, \bibinfo {author} {\bibfnamefont {Deborah~G.}\ \bibnamefont
  {Evans}}, \ and\ \bibinfo {author} {\bibfnamefont {Abraham}\ \bibnamefont
  {Nitzan}},\ }\bibfield  {title} {\enquote {\bibinfo {title} {{A
  nonequilibrium golden rule formula for electronic state populations in
  nonadiabatically coupled systems}},}\ }\href {\doibase 10.1063/1.468153}
  {\bibfield  {journal} {\bibinfo  {journal} {J. Chem. Phys.}\ }\textbf
  {\bibinfo {volume} {101}},\ \bibinfo {pages} {436} (\bibinfo {year}
  {1994})}\BibitemShut {NoStop}%
\end{thebibliography}%

\end{document}